\documentclass[useAMS,usenatbib]{mnras}	
\usepackage[figuresright]{rotating}
\usepackage{pdflscape} 

\usepackage{graphics}

\usepackage{bigdelim}
\usepackage{bigstrut}

\usepackage{setspace}

\usepackage{natbib}
\usepackage{mathtools}
\usepackage{amsmath}
\usepackage{amssymb}
\usepackage[subnum]{cases}

\title[WAGGS]{The WAGGS project - I. The WiFeS Atlas of Galactic Globular cluster Spectra}

\author[Usher et al.]	
  {Christopher~Usher$^{1, 2}$\thanks{email: C.G.Usher@ljmu.ac.uk},
  {Nicola~Pastorello$^{3, 2}$},
  {Sabine~Bellstedt$^{2}$},
  {Adebusola~Alabi$^{2}$},\newauthor
  {Pierluigi~Cerulo$^{2,4}$},
  {Leonie Chevalier$^{2}$},
  {Amelia~Fraser-McKelvie$^{5,6}$},\newauthor
  {Samantha~Penny$^{7}$},
  {Caroline~Foster$^{8}$},
  {Richard~M.~McDermid$^{8,9}$},\newauthor
  {Ricardo~P.~Schiavon$^{1}$},
  {Alexa~Villaume$^{10}$}
  \\
   $^1$Astrophysics Research Institute, Liverpool John Moores University, 146 Brownlow Hill, Liverpool L3 5RF, UK
  \\ $^2$Centre for Astrophysics \& Supercomputing, Swinburne University of Technology, Hawthorn, VIC 3122, Australia
  \\ $^{3}$Deakin Software and Technology Innovation Laboratory, Deakin University, Burwood, VIC 3125, Australia
  \\ $^4$Departamento de Astronomia, Universidad de Concepcion, Casilla 160-C, Chile
  \\ $^5$School of Physics \& Astronomy, Monash University, Clayton, VIC 3800, Australia 
  \\ $^6$Monash Centre for Astrophysics (MoCA), Monash University, Clayton, VIC 3800, Australia
  \\ $^7$Institute of Cosmology and Gravitation, University of Portsmouth, Dennis Sciama Building, Burnaby Road, Portsmouth PO1 3FX, UK
  \\ $^8$Australian Astronomical Observatory, PO Box 915, North Ryde, NSW 1670, Australia
  \\ $^9$Department of Physics and Astronomy, Macquarie University, North Ryde NSW 2109, Australia
  \\ $^{10}$University of California Observatories, 1156 High Street, Santa Cruz, CA 95064, USA
}
\begin{document}

\maketitle

\begin{abstract}
We present the WiFeS Atlas of Galactic Globular cluster Spectra, a library of integrated spectra of Milky Way and Local Group globular clusters.
We used the WiFeS integral field spectrograph on the Australian National University 2.3 m telescope to observe the central regions of 64 Milky Way globular clusters and 22 globular clusters hosted by the Milky Way's low mass satellite galaxies.  
The spectra have wider wavelength coverage (3300 \AA{} to 9050 \AA{}) and higher spectral resolution ($R = 6800$) than existing spectral libraries of Milky Way globular clusters.
By including Large and Small Magellanic Cloud star clusters, we extend the coverage of parameter space of existing libraries towards young and intermediate ages.
While testing stellar population synthesis models and analysis techniques is the main aim of this library, the observations may also further our understanding of the stellar populations of Local Group globular clusters and make possible the direct comparison of extragalactic globular cluster integrated light observations with well understood globular clusters in the Milky Way.
The integrated spectra are publicly available via the project website.
\end{abstract}

\begin{keywords}
globular clusters: general, Local Group, galaxies: abundances, galaxies: stellar content 
\end{keywords}

\section{Introduction}
\label{sec:introduction}
To measure the age, chemical composition or initial mass function (IMF) of a stellar population from its integrated light requires reliable stellar population synthesis models.
Globular clusters (GCs) provide an important laboratory for testing stellar population synthesis models due to their relatively simple stellar populations.
Therefore, the task of studying their stellar populations is much simpler than for galaxies which contain stars with a wide range of ages and metallicities. 
In the case of Milky Way (MW) GCs, detailed chemical abundances are available from high resolution spectroscopy of individual stars \citep[e.g.][]{2009A&A...508..695C} while ages \citep[e.g.][]{2009ApJ...694.1498M, 2013ApJ...775..134V} and mass functions \citep[e.g.][]{2010AJ....139..476P} are available from resolved imaging. 
By comparing the stellar population parameters derived from the integrated star light with the parameters measured from individual stars, we can test the reliability of stellar population synthesis models.

Additionally, GCs can be used to gain important insights into how galaxies form and evolve.
There is a long history of using GCs to study both our own galaxy \citep[e.g.][]{1918PASP...30...42S, 1978ApJ...225..357S} and external galaxies \mbox{\citep[e.g.][]{1977MNRAS.180..309H, 1984ApJ...287..586B, 1991ApJ...379..157B, 1998ApJ...496..808C}}.
Due to their high surface brightnesses, the integrated light of GCs can be studied spectroscopically at much greater distances than individual stars.
For example, \citet{2005A&A...439..997P} studied the stellar populations of GCs around NGC 7192 ($D = 38$ Mpc) and \citet{2011A&A...531A...4M} studied the kinematics of GCs in the Hydra I galaxy cluster ($D = 47$ Mpc).
This allows GCs to be used to perform stellar archaeology beyond the Local Group.
Integrated light observations of MW GCs can be directly compared with observations of extragalactic GCs in a model independent way.
Large datasets of extragalactic GC spectra are now available with the SAGES Legacy Unifying Globulars and GalaxieS \citep[SLUGGS, ][]{2014ApJ...796...52B, sluggs_rv} survey providing spectra of over 4000 GCs in 27 galaxies.
Other large datasets are available for a few individual galaxies with spectra observed of 922 GCs around M87 \citep{2011ApJS..197...33S, 2014ApJ...792...59Z}, of over 693 GCs around NGC 1399 \citep{2010A&A...513A..52S} and of 563 GCs around NGC 5128 \citep{2010AJ....139.1871W}.

Although they typically do not show spreads in age or iron abundance \citep[e.g.][]{2009A&A...508..695C}, GCs show star-to-star abundance variations in elements such as helium, oxygen and sodium \citep[e.g.][]{2012A&ARv..20...50G}.
These variations are often explained as GCs consisting of multiple generations of stars, with the later generation(s) forming out of material enriched by the products of hot hydrogen burning.
Even though several scenarios for the formation of multiple generations have been proposed \citep[e.g.][]{2007A&A...464.1029D, 2009A&A...507L...1D, 2010MNRAS.407..854D, 2014MNRAS.437L..21D}, none are consistent with observations \citep{2014MNRAS.443.3594B, 2015MNRAS.449.3333B, 2015MNRAS.453..357B}.
Due to the effects of dynamical evolution, the mass functions of GCs today are usually different \citep[e.g.][]{2007ApJ...656L..65D, 2010AJ....139..476P} than what would be expected from the effects of stellar evolution on a `normal' IMF.
While these properties make GCs less than perfect examples of simple stellar populations, reliable comparisons with stellar population synthesis models are still possible if their mass functions and chemical abundance distributions are well measured.
Extragalactic GCs are likely similarly affected by multiple populations and dynamical effects, so comparing extragalactic GC observations to those of MW GCs can be preferable to using models (although the mean stellar populations of MW GCs may not be typical of GCs in all galaxies).

Several studies, including \citet{1984ApJ...287..586B}, \citet{1984ApJS...55...45Z}, \citet{1986ApJ...300..258B}, \citet{1986A&A...162...21B}, \citet{1988AJ.....96...92A}, \citet{1995A&A...303...79C}, \citet{1998ApJ...496..808C}, \citet{2002MNRAS.336..168B}, \citet{2002A&A...395...45P}, \citet{2005ApJS..160..163S} and \citet{2011A&A...530A..22P}, have obtained integrated low resolution spectra of MW, Large Magellanic Cloud (LMC) and Small Magellanic Cloud (SMC) GCs with the aim of stellar population analysis.
With the exceptions of \citet[$R \sim 2000$]{1988AJ.....96...92A} and \citet[$R \sim 1600$]{2005ApJS..160..163S} these studies have been at the low ($R \sim 600$) spectral resolution typical of the Lick system \citep{1994ApJS...95..107W, 2007ApJS..171..146S}.
In addition, a few \citep[e.g.][]{2011ApJ...735...55C, 2012A&A...546A..53L, 2013MNRAS.434..358S, 2017ApJ...834..105C, 2017arXiv170207353L} studies have obtained integrated high resolution ($R \sim 20000$) spectra to measure detailed chemical abundances of MW and MW satellite GCs; however these studies suffer from small sample sizes ($< 10$ GCs each). 
Most of these studies, except for \citet{2005ApJS..160..163S}, have only made their spectral index measurements public and not the spectra themselves.
Publicly available spectra make possible the comparison of full observed spectral energy distribution with models and enables the testing of novel analysis techniques.

Recently, interest has been growing in using the redder (redwards of H$\alpha$) regions of the optical wavelength range for stellar population studies, many of which have focussed on the calcium triplet (CaT) at 8498, 8542 and 8662 \AA{}.
First studied as a metallicity indicator in integrated GC spectra by \citet{1988AJ.....96...92A}, the CaT has been used as an age-insensitive method to estimate metallicities of extragalactic GCs \citep{2010AJ....139.1566F, 2012MNRAS.426.1475U, 2015MNRAS.451.2625P} and integrated galaxy light \citep{2009MNRAS.400.2135F, 2014MNRAS.442.1003P}.
Intriguingly, the relationship between CaT strength and GC colour has been observed to vary between galaxies \citep{2011MNRAS.415.3393F, 2012MNRAS.426.1475U, 2015MNRAS.446..369U} which could be caused by variations in the colour--metallicity relation or by systematics related to the CaT.
Additionally, the reliability of the CaT as a metallicity indicator at high metallicity has been challenged \citep{2016ApJ...818..201C}.

The redder part of the optical region is also of interest to studies of the IMF as it contains spectral features including the CaT and the sodium doublet at 8190 \AA{} which are sensitive to the ratio of dwarf to giant stars \citep[e.g.][]{1997ApJ...479..902S, 2000ApJ...532..453S}.
While multiple studies \citep[e.g.][]{2012ApJ...760...71C, 2013MNRAS.429L..15F} have claimed that the slope of the IMF increases with stellar mass or other galaxy properties, it is unclear if the stellar population synthesis models used correctly account for abundance variations \citep[e.g.][]{2015MNRAS.454L..71S}. 

Unfortunately, there is a lack of high quality MW GC integrated spectra covering the red optical wavelengths. 
The best existing spectral library of MW GCs, \citet{2005ApJS..160..163S}, only covers the wavelength range from 3360 \AA{} to 6430 \AA{} which does not include lines such as the CaT and the 8190 \AA{} sodium doublet that are utilised in IMF studies.
Furthermore, the spectral resolution of Schiavon et al. is low ($R \sim 1600$) compared to those now available for large numbers of extragalactic GCs \citep[e.g. $R \sim 5000$ for][]{2012MNRAS.426.1475U}. 
Higher spectral resolution data makes possible the deblending of spectral lines and more detailed chemical abundances to be measured. 
Higher resolution spectra can always be degraded for comparison with low resolution spectra; the reverse is not true.

While the GC system of the MW is relatively well studied, it provides only a limited range of ages and chemical compositions.
As different galaxies have different evolutionary histories, this will be reflected in variations in their GC properties such as ages, metallicities, and abundance patterns.
However, the proximity of other Local Group galaxies allows us to extend this parameter space. 
As would be expected for galaxies spanning a wide range in stellar mass, the Fornax dwarf spheroidal (Fornax), LMC, SMC and the MW all have different age--metallicity and [$\alpha$/Fe]--[Fe/H] relationships\citep[e.g.][]{2009AJ....138.1243H, 2012A&A...544A..73D, 2013AJ....145...17P, 2013A&A...560A..44V, 2015ApJ...806...21D}.
Both SMC and LMC host massive young and intermediate age star clusters which are rare and difficult to study in the MW  \citep[e.g. GLIMPSE-CO1][]{2005AJ....129..239K, 2011MNRAS.411.1386D}.

To provide a library of integrated spectra of the GCs of the MW and its satellite galaxies with wider wavelength coverage and higher spectral resolution, we have turned to integral field spectroscopy to produce WAGGS - the WiFeS Atlas of Galactic Globular cluster Spectra.
Using the Wide-Field Spectrograph (WiFeS) on the Australian National University (ANU) 2.3 metre telescope we have observed 64 GCs in the MW, 3 in the Fornax dSph, 14 in the LMC and 5 in the SMC.
Besides providing spatially-resolved spectroscopy at higher resolution ($R \sim 6800$) and wider wavelength coverage (3270 \AA{} to 9050 \AA{}) than existing spectral libraries, WAGGS includes more GCs (86) and a wider range of ages (20 Myr to 13 Gyr).

This paper is the first in a series of papers based on the WAGGS data.
In this paper we describe our sample selection and the properties of our sample in Section \ref{sec:sample}.
In Section \ref{sec:observations} we describe our observations and our data reduction methods.
Finally, in Section \ref{sec:summary} we discuss planned and possible uses of the WAGGS spectra.

Throughout this paper we will use the term GC to refer to all observed star clusters even though a number of the massive ($> 10^{4}$ M$_{\sun}$) star clusters observed in the SMC and LMC are significantly younger than classical GCs.
We note that the definition of a GC has been debated in the literature \citep{2011PASA...28...77F, 2012AJ....144...76W}.
With the possible exception of the very youngest ($\sim 20$ Myr) LMC and SMC star clusters, all of our GCs meet the definition of \citet{2015MNRAS.454.1658K}, namely ``[a] gravitationally bound, stellar cluster that in terms of its position and velocity vectors does not coincide with the presently star-forming component of its host galaxy''.
We note that two objects included in our sample, NGC 5139 \citep[$\omega$ Cen, e.g.][]{2010ApJ...722.1373J} and NGC 6715 \citep[M 54, e.g. ][]{2010A&A...520A..95C}, show significant metallicity spreads and are thought to have formed from the centres of accreted dwarf galaxies \citep[e.g.][]{2000A&A...362..895H, 1997AJ....113..634I}.
As such they may be more akin to ultra-compact dwarfs \citep{2014MNRAS.444.3670P}.

\section{Sample selection and properties}
\label{sec:sample}
Our aim in selecting GCs for observation was to obtain a representative sample of GCs rather than a complete one.
The starting point for our sample is the Advanced Camera for Surveys (ACS) Globular Cluster Treasury Survey \citep{2007AJ....133.1658S} which provides high quality, homogeneous Hubble Space Telescope (HST) ACS imaging for 65 Galactic GCs.
We supplemented these with the 6 MW halo GCs observed with ACS in a similar manner by \citet{2011ApJ...738...74D}.
Many of these clusters also have UV and blue HST Wide Field Camera 3 (WFC3) photometry from the HST UV Legacy Survey of Galactic Globular Clusters \citep{2015AJ....149...91P}. 
High quality age measurements \citep[e.g.][]{2009ApJ...694.1498M, 2013ApJ...775..134V} and horizontal branch morphologies \citep{2014ApJ...785...21M} are available for the majority of these GCs.
We restricted our sample to GCs with declinations lower than $+30^{\circ}$ so that they would be observable from Siding Spring Observatory and central surface brightness brighter than $\mu_{V} \sim 20$ mag arcsec$^{-2}$ so that we could obtain high signal-to-noise ratios (S/N) in reasonable exposure times.
We prioritised GCs which already have detailed chemical abundance measurements available for multiple stars such as the GCs observed by \citet{2009A&A...508..695C}, by ESO-Gaia \citep{2012Msngr.147...25G} and by APOGEE \citep{2016AN....337..863M, 2015AJ....149..153M}. 
We include NGC 5139 and NGC 6715 to test how well the mean properties of complex stellar populations can be measured from integrated light.

In addition, we observed a number of bright MW GCs not in the ACS Globular Cluster Treasury Survey that have published chemical abundances from high resolution spectroscopy, many of which have HST Wide Field and Planetary Camera 2 photometry from \citet{2002A&A...391..945P} and relative ages from \citet{2005AJ....130..116D}.
We also observed a number of bright but relatively poorly studied GCs, including NGC 5824, NGC 6284, NGC 6316, NGC 6333 and NGC 6356, both to improve our coverage of bulge and halo GCs and to provide a larger sample of GCs for integrated light abundance studies.

To expand the range of ages and chemical compositions observed, we supplemented our sample with GCs in MW satellites.
We targeted three old (age $> 10$ Gyr) GCs in the Fornax dSph, one in the SMC and four in the LMC.
We also observed eight intermediate age ($1 <$ age $< 10$ Gyr) and six young (age $< 1$ Gyr) GCs in the LMC and SMC.
Two of our MW GCs, NGC 5634 and NGC 6715 can be associated with the disrupting Sagittarius dwarf spheroidal galaxy on the basis of their positions and kinematics \citep[e.g.][]{2003AJ....125..188B, 2010ApJ...718.1128L} while a third (NGC 4147) shows only weak evidence of being associated with the Sagittarius dwarf \citep{2010ApJ...718.1128L}.
As with our MW targets, we prioritised bright GCs with HST imaging and high resolution spectroscopy.

\begin{figure*}
\begin{center}
\includegraphics[width=504pt]{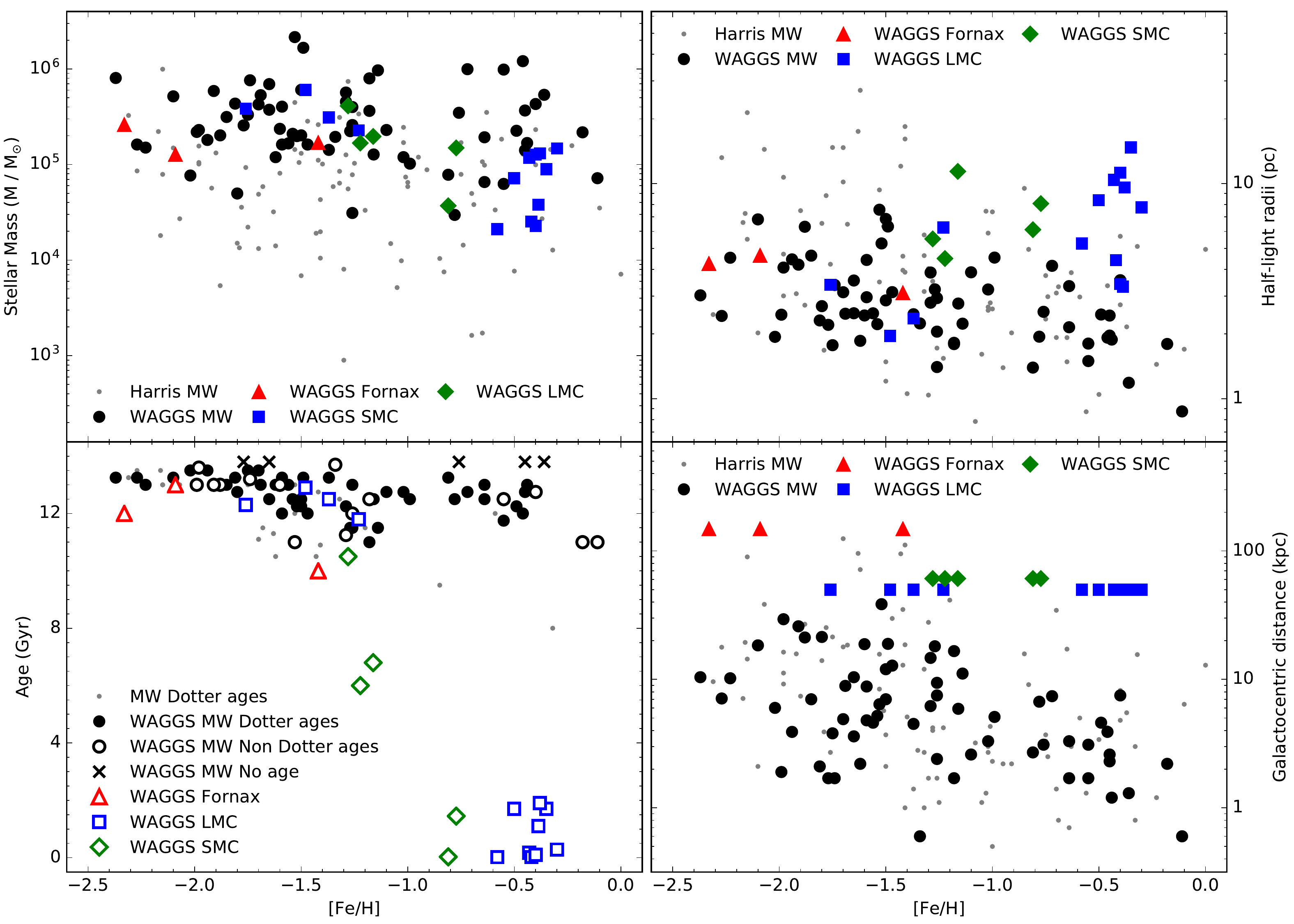}
\caption{General properties of the WAGGS sample.
\emph{Top left:} Stellar mass from $V$-band luminosity versus metallicity.
WAGGS GCs in the MW are plotted as black circles, WAGGS GCs in Fornax as red triangles, WAGGS GCs in the LMC as blue squares and WAGGS GCs in the SMC as green diamonds.
Grey points are MW GCs from the 2010 edition of the \citet{1996AJ....112.1487H} catalogue.
The WAGGS sample is biased towards higher masses relative to the MW population.  
\emph{Bottom left:} Age versus metallicity.
The filled WAGGS points have ages from \citet{2010ApJ...708..698D}, \citet{2011ApJ...738...74D} or \citet{2014ApJ...785...21M}; open points have ages from other sources.
MW GCs marked with crosses have no published ages and have been assigned ages of 13.8 Gyr for the proposes of illustration.
The grey points are other MW GCs with ages from \citet{2010ApJ...708..698D}, \citet{2011ApJ...738...74D} or \citet{2014ApJ...785...21M}.
The WAGGS sample covers much of the MW GCs age-metallicity space as well as the range of GC ages in the LMC and SMC.
\emph{Top right:} Half-light radius versus metallicity.
The WAGGS GCs are plotted as in the top left panel and the grey points are all MW GCs in the Harris catalogue.
\emph{Bottom right:} Galactocentric distance versus metallicity.
The WAGGS GCs and the Harris MW GCs are plotted as in the top right panel.
The WAGGS sample misses faint, extended GCs found in the MW halo as well as heavily extincted GCs in the bulge.
}
\label{fig:sample_properties}
\end{center}
\end{figure*}

An overview of the properties of the sample GCs is shown in Figure \ref{fig:sample_properties} and given in Table \ref{tab:sample_properties}.
For the MW GCs we use metallicities, distances and structural parameters from the 2010 edition of the \citet{1996AJ....112.1487H} catalogue.
The Harris metallicities are on the \citet{2009A&A...508..695C} scale but come from a range of sources. 
For MW GCs, we prefer ages estimated by \citet{2010ApJ...708..698D} and \citet{2011ApJ...738...74D}, supplemented by ages estimated by \citet{2014ApJ...785...21M} using the same techniques and models.
Together, these are the largest homogeneous sample of MW GC ages.
Other sources of ages are given in Table \ref{tab:sample_properties}.
We scaled the \citet{2005AJ....130..116D} and \citet{2006A&A...456.1085M} ages to match the Dotter et al. measurements by using GCs of similar ages and metallicities in both studies.
\citet{2008ApJ...687L..79O} claim that NGC 6440 has the same age as NGC 104; therefore we assign NGC 6440 the same age as calculated by \citet[12.75 Gyr]{2010ApJ...708..698D} for NGC 104.

For Fornax, the LMC and the SMC we adopted the distances used by \citet{2005ApJS..161..304M}: 137 kpc, 50 kpc and 60 kpc respectively.
We drew the structural parameters of GCs around these galaxies from several studies listed in Table \ref{tab:sample_properties}, using the values provided by \citet{2005ApJS..161..304M} when available.
Likewise, ages and metallicities for the GCs of MW satellites come from a range of studies listed in Table \ref{tab:sample_properties}.
For the satellite galaxies, we preferred metallicities from high resolution spectroscopy to those from the resolved star CaT strengths, and CaT based metallicities to those based on resolved colour-magnitude diagrams.
For the LMC and SMC clusters with metallicities based on the strength of the CaT, we recalculated the metallicities using CaT-metallicity relations based on the \citet{2009A&A...508..695C} metallicity scale.
For NGC 416, NGC 419 and NGC 1846, for which metallicities were derived using equation 5 of \citet{2004MNRAS.347..367C} by \citet{2009AJ....138.1403G} or \citet{2006AJ....132.1630G}, we recalculated the metallicities using equation 6 of \citet{2004MNRAS.347..367C}.
For NGC 1868, we converted the metallicity of \citet{1991AJ....101..515O} from the \citet{1984ApJS...55...45Z} scale to the \mbox{\citet{2009A&A...508..695C}} scale using the equation provided by \citet{2009A&A...508..695C}.

We estimated stellar masses for each of our GCs using their extinction corrected $V$-band absolute magnitudes.
For each GC, we used the extinction measurements provided by the same source that provided the structural parameters (the Harris catalogue for MW GCs, \citealt{2005ApJS..161..304M} for most others).
Stellar population synthesis models \citep[e.g.][]{2009ApJ...699..486C} predict that the $V$-band mass-to-light ratio ($M/L_{V}$) increases with metallicity, while dynamical studies \citep[e.g.][]{2011AJ....142....8S, 2017MNRAS.464.2174B} show that the $M/L_{V}$ decreases with metallicity or remains constant.
For this reason we have adopted the agnostic approach of a constant $M/L_{V} = 2$ for all GCs older than 10 Gyr.
For younger GCs, we follow \citet{2005ApJS..161..304M} and use the \citet{2003MNRAS.344.1000B} derived values.
We use luminosity based masses as dynamical masses are not available for all GCs in our sample.
The properties shown in Figure \ref{fig:sample_properties} and given in Table \ref{tab:sample_properties} highlight the range of sample properties but may not be the final values adopted for future analysis.

In general, we favoured GCs with higher central surface brightnesses in order to maximise the number of GCs observed in the available observing time.
As such, our sample suffers from a few biases.
First, our sample is biased towards more massive GCs (see the upper left panel of Figure \ref{fig:sample_properties}).
We note that extragalactic GC studies are usually limited to the mean of the globular cluster luminosity function, if not brighter.
Massive GCs would also be better for stellar population comparisons as the effects of stochastic sampling of the IMF would be less severe.
Second, our sample is biased towards spatially concentrated GCs (see the upper right panel of Figure \ref{fig:sample_properties}).
As such, we miss many of the faint, extended GCs in the MW halo.
Third, our sample is biased against GCs with high foreground reddenings.
While this biases our sample against bulge GCs, these highly reddened GCs are typically poorly studied, thus making them poor choices for testing stellar population synthesis models.
These selection biases are mostly shared with previous integrated light studies.
Our sample includes 35 of the 40 GCs observed by \citet{2005ApJS..160..163S} and 10 of the 12 GC observed by \citet{2002A&A...395...45P}.
Compared to \citet{2005ApJS..160..163S}  and \citet{2002A&A...395...45P} our sample extends to lower mass MW GCs and includes more metal poor GCs and halo GCs.

\begin{figure*}
\begin{center}
\includegraphics[width=504pt]{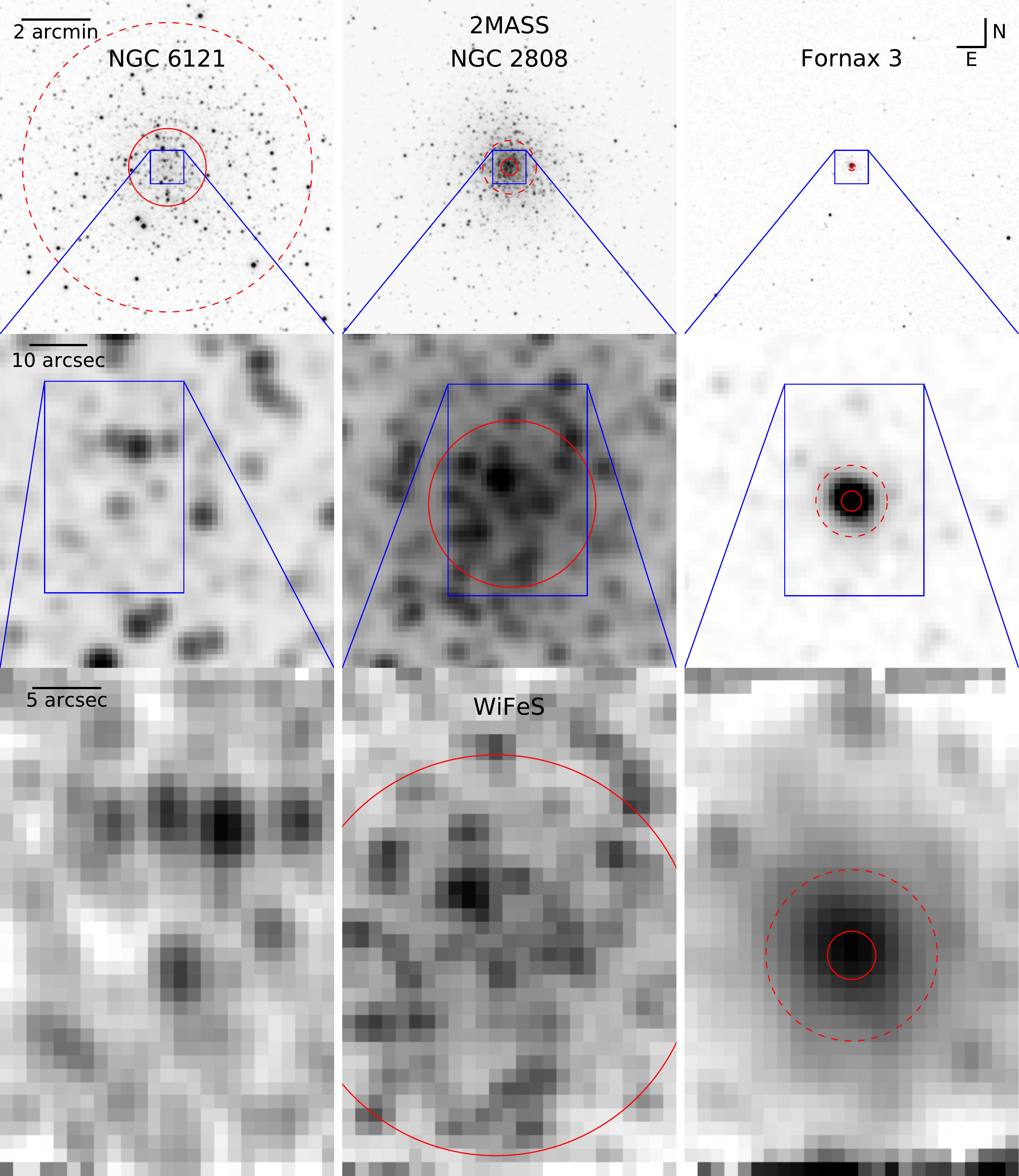}
\caption{Field-of-view of WAGGS observations.
In the top row are 2MASS images for NGC 6121 (M 4, left), NGC 2808 (centre) and Fornax 3 (NGC 1049, right).
Each image is 10 by 10 arcmin with North up and East left.
In middle row, 1 by 1 arcmin 2MASS images are shown for the same GCs.
The footprints of these panels are shown as blue squares in the upper row.
In the lower row the WAGGS I7000 datacubes summed along the spectral direction are shown.
The 25 by 38 arcsec footprints of these datacubes are shown as blue rectangles in the middle row.
In each panel, the solid red curve shows the core radius while dashed red circle shows the half-light radius.
The three GCs pictured span the range of angular sizes in the sample with NGC 6121 being the most extended in our sample, NGC 2808 being close to the median and Fornax 3 being the least extended GC.
We note that many of the `stars' in both the 2MASS imaging and the WAGGS cubes are blends of multiple stars.}
\label{fig:field_of_view}
\end{center}
\end{figure*}

\begin{figure}
\begin{center}
\includegraphics[width=240pt]{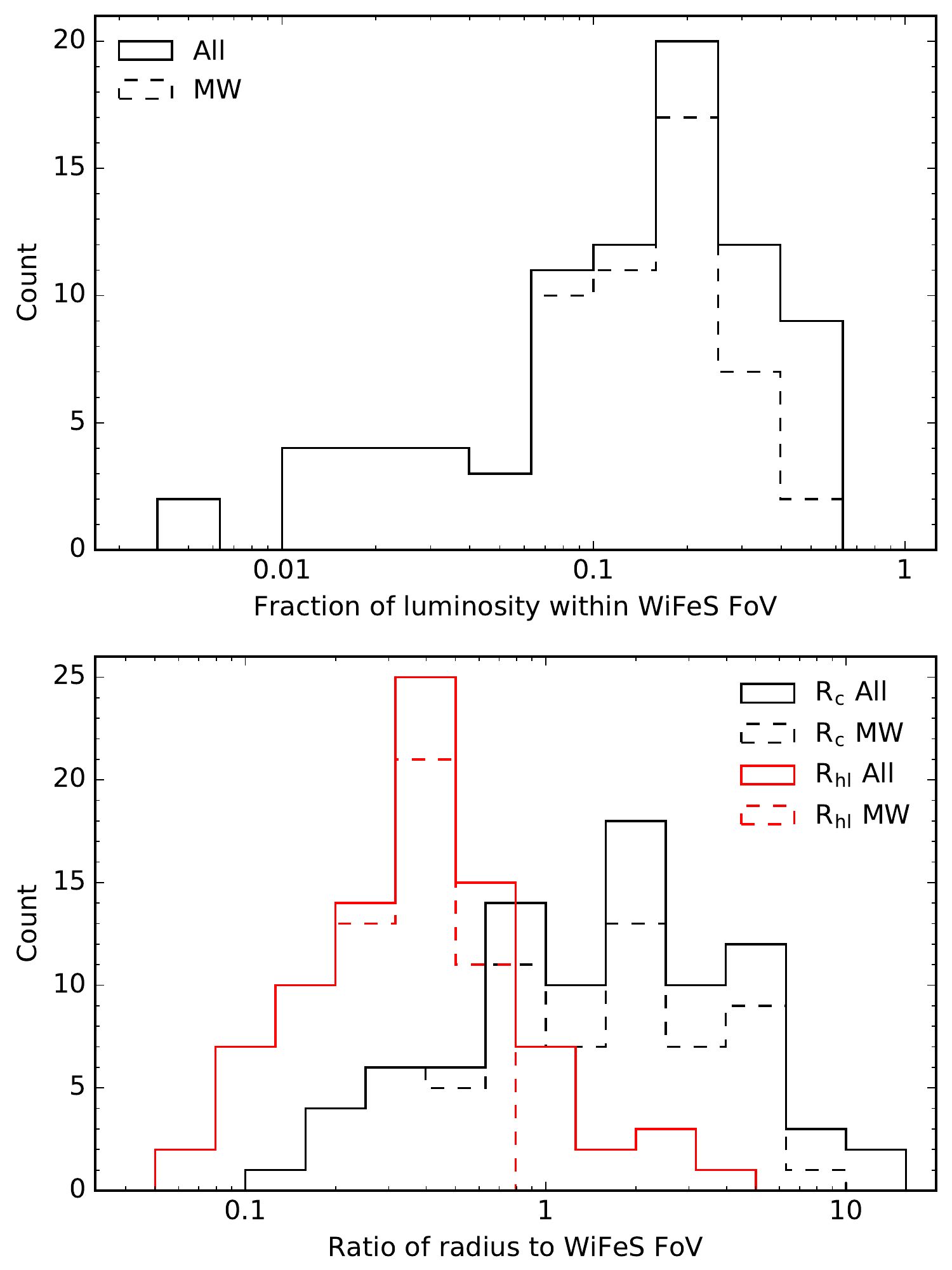}
\caption{Spatial extent of the WAGGS observations.
\emph{Top:} Fraction of GC $V$-band luminosity within the WiFeS field-of-view.
The solid line shows all WAGGS GCs while the dashed line shows only MW GCs.
The luminosity within the field-of-view was calculated by integrating the surface brightness profiles calculated from the structural parameters given in Table \ref{tab:sample_properties}.
Typically we observe a fraction of 0.19 of the total luminosity of a GC (0.12 in the Milky Way).
\emph{Bottom:} Ratio of core radius (black) and half-light radius (red) to the mean radius of the WiFeS field-of-view (17.4 arcsec).
The solid lines shows all WAGGS GCs while the dashed lines shows only MW GCs.
Typically we observe out to 1.7 times the core radius and 0.37 times the half-light radius (1.3 and 0.32 times respectively for MW GCs).}
\label{fig:fov_fractions}
\end{center}
\end{figure}

Since we only observed a single, central pointing for each GC and the GCs in our sample span a wide range of heliocentric distances (from 2.2 kpc for NGC 6121 to 137 kpc for the Fornax GCs), the fraction of GC light we observed varies dramatically.
This is illustrated in Figure \ref{fig:field_of_view} where we show both the nearest (NGC 6121) and farthest (Fornax 3) GCs in sample as well a GC at the median distance (10 kpc, NGC 2808).
The fraction of GC $V$-band light within the WiFeS field-of-view was estimated from surface brightness profiles calculated from the structural parameters given in Table \ref{tab:sample_properties} using the \textsc{limepy} code \citep{2015MNRAS.454..576G} and a \citet{1966AJ.....71...64K} profile.
The fraction of observed luminosity ranges from 0.005 (NGC 5139) to 0.81 (Fornax 4) with a median of 0.17 (0.12 for MW GCs).
These observed light fractions are shown in the top panel of Figure \ref{fig:fov_fractions}.
In the bottom panel of Figure \ref{fig:fov_fractions}, we also show the ratios of the core radius and half-light radius to the equivalent radius of the WiFeS field-of-view (17.4 arcsec). 
Our observations extend to between 0.12 (NGC 5139) and 13 (Fornax 5) core radii with a median of 1.7 (1.3 for MW GCs) and to between 0.06 (NGC 5139) and 3.7 (Fornax 4) half-light radii with a median of 0.37 (0.32 for MW GCs).
We note that previous integrated light studies of MW GCs \citep[e.g.][]{2002A&A...395...45P, 2005ApJS..160..163S, 2017ApJ...834..105C} were also typically limited to within a core radius.

The aperture bias affecting the WAGGS spectra is a concern for a few reasons.
First, in the case where the WiFeS field-of-view only covers a small fraction of a GC's area, stochastic effects could prevent the proper sampling of all stages of stellar evolution. 
This can be seen in the images of NGC 6121 in Figure \ref{fig:field_of_view} and is especially a concern for lower-mass clusters.
Second, GCs are affected by mass segregation as more massive stars sink to their cluster centres due to dynamical evolution.
This causes the slope of the mass function to vary with radius \citep[e.g.][]{1999ApJ...523..752R, 2003AJ....126..815L, 2004A&A...425..509A, 2015ApJ...814..144B}, which mimics the effect of the IMF varying weakly with radius.
Third, the ratio of first to second generation stars is observed to vary with radius in some GCs \citep[e.g.][]{2011A&A...525A.114L, 2015ApJ...804...71L}.
This creates radial gradients in the mean chemical abundances.
The latter two effects can be accounted for in comparisons with stellar population synthesis models if the mass function and the chemical abundance distributions are known for the same region as the WiFeS field-of-view, but remain an issue for comparisons with extragalactic GCs. 
We aim to explore the importance of these effects with our data in future works.

While our sample is subject to observational biases, it is the largest yet sample of integrated spectra of MW GCs.
The sample spans the full range of MW GCs metallicities ($-2.4 <$ [Fe/H] $< -0.1$) and a wide range of ages (20 Myr to 13.5 Gyr).
The samples includes MW GCs from within a kpc of the Galactic centre (NGC 6528) to nearly 40 kpc out in the MW halo (NGC 7006) as well as 22 GCs in MW satellites.

\addtolength{\tabcolsep}{-2pt}

\begin{landscape}
\begin{table}
\caption{General properties of the WAGGS sample}
\label{tab:sample_properties}
\begin{tabular}{cccccccccccccccccc}
\hline
ID & Name & Galaxy & RA & Dec & $r_{\sun}$ & $r_{gc}$ & [Fe/H] & Age & $R_{c}$ & $R_{h}$ & $\mu_{V0}$ & $A_{V}$ & GC  & FoV  & [Fe/H] & Age & Structural \\
& & & & & & & & & & & & & Mass & Mass & Source & Source & Source \\
& & & [$^{\circ}$] & [$^{\circ}$] & [kpc] & [kpc] & [dex] & [Gyr] & [arcsec] & [arcsec] & [mag arcsec$^{-2}$] & [mag] & [$\log$ M$_{\sun}$]  & [$\log$ M$_{\sun}$] &  &  &   \\
(1) & (2) & (3) & (4) & (5) & (6) & (7) & (8) & (9) & (10) & (11) & (12) & (13) & (14) & (15) & (16) & (17) & (18) \\ \hline
NGC 104 & 47 Tuc & MW & 6.024 & $-$72.081 & 4.5 & 7.4 & $-$0.72 & 12.8 & 22 & 190 & 14.4 & 0.12 & 6.0 & 4.7 & H & Do10 & H \\
NGC 121 &  & SMC & 6.701 & $-$71.536 & 60.0 & 61.0 & $-$1.28 & 10.5 & 10 & 19 & 18.3 & 0.45 & 5.6 & 5.2 & Da16 & Gl08a & McL05 \\
NGC 330 &  & SMC & 14.074 & $-$72.463 & 60.0 & 61.0 & $-$0.81 & 0.03 & 8 & 21 & 16.5 & 0.20 & 4.6 & 4.2 & Hi99 & Si02 & McL05 \\
NGC 361 &  & SMC & 15.542 & $-$71.605 & 60.0 & 61.0 & $-$1.16 & 6.8 & 24 & 39 & 20.2 & 0.21 & 5.3 & 4.5 & Da98 & Mi98 & McL05 \\
NGC 362 &  & MW & 15.809 & $-$70.849 & 8.6 & 9.4 & $-$1.26 & 11.5 & 11 & 49 & 14.8 & 0.15 & 5.6 & 4.9 & H & Do10 & H \\
NGC 416 &  & SMC & 16.994 & $-$72.355 & 60.0 & 61.0 & $-$1.22 & 6.0 & 10 & 15 & 18.5 & 0.39 & 5.2 & 4.9 & Gl09 & Gl08b & McL05 \\
NGC 419 &  & SMC & 17.072 & $-$72.883 & 60.0 & 61.0 & $-$0.77 & 1.4 & 13 & 28 & 17.8 & 0.31 & 5.2 & 4.9 & Gl09 & Go14 & Gl09 \\
Fornax 3 & NGC 1049 & Fornax & 39.951 & $-$34.258 & 137.0 & 149.0 & $-$2.33 & 12.0 & 2 & 6 & 17.1 & 0.11 & 5.4 & 5.3 & La12 & deB16 & McL05 \\
Fornax 4 &  & Fornax & 40.032 & $-$34.536 & 137.0 & 149.0 & $-$1.42 & 10.0 & 2 & 5 & 17.5 & 0.43 & 5.2 & 5.1 & La12 & deB16 & McL05 \\
Fornax 5 &  & Fornax & 40.588 & $-$34.102 & 137.0 & 149.0 & $-$2.09 & 13.0 & 1 & 7 & 17.6 & 0.10 & 5.1 & 5.0 & La12 & deB16 & McL05 \\
NGC 1261 &  & MW & 48.068 & $-$55.216 & 16.3 & 18.1 & $-$1.27 & 11.5 & 21 & 41 & 17.7 & 0.03 & 5.3 & 4.4 & H & Do10 & H \\
NGC 1786 &  & LMC & 74.783 & $-$67.745 & 50.1 & 50.0 & $-$1.76 & 12.3 & 4 & 14 & 16.4 & 0.39 & 5.6 & 5.3 & Mu10 & Ge97 & McL05 \\
NGC 1783 &  & LMC & 74.787 & $-$65.987 & 50.1 & 50.0 & $-$0.35 & 1.7 & 38 & 61 & 20.4 & 0.02 & 5.0 & 3.9 & Mu08 & Go14 & Go06+11 \\
NGC 1846 &  & LMC & 76.891 & $-$67.461 & 50.1 & 50.0 & $-$0.50 & 1.7 & 26 & 34 & 19.6 & 0.08 & 4.9 & 4.1 & Gr06 & Go14 & Go06+09 \\
NGC 1850 &  & LMC & 77.186 & $-$68.762 & 50.1 & 50.0 & $-$0.40 & 0.1 & 11 & 46 & 16.7 & 0.32 & 5.1 & 4.4 & Ni15 & Ni15 & McL05 \\
NGC 1856 &  & LMC & 77.372 & $-$69.128 & 50.1 & 50.0 & $-$0.30 & 0.3 & 7 & 32 & 16.8 & 0.71 & 5.2 & 4.6 & Ba13 & Ba13 & McL05 \\
NGC 1866 &  & LMC & 78.413 & $-$65.466 & 50.1 & 50.0 & $-$0.43 & 0.2 & 11 & 43 & 17.3 & 0.40 & 5.1 & 4.4 & Mu11 & Ba13 & McL05 \\
NGC 1851 &  & MW & 78.528 & $-$40.047 & 12.1 & 16.6 & $-$1.18 & 11.0 & 5 & 31 & 14.2 & 0.06 & 5.6 & 5.0 & H & Mi14 & H \\
NGC 1868 &  & LMC & 78.651 & $-$63.955 & 50.1 & 50.0 & $-$0.39 & 1.1 & 6 & 14 & 17.8 & 0.41 & 4.6 & 4.3 & Ol91 & Ke07 & McL05 \\
NGC 1898 &  & LMC & 79.177 & $-$69.656 & 50.1 & 50.0 & $-$1.23 & 11.8 & 9 & 26 & 18.6 & 0.26 & 5.4 & 4.9 & Jo06 & Ol98 & McL05 \\
NGC 1916 &  & LMC & 79.652 & $-$69.407 & 50.1 & 50.0 & $-$1.48 & 12.9 & 3 & 8 & 15.3 & 0.58 & 5.8 & 5.6 & Co11 & Ma03 & McL05 \\
NGC 1904 & M 79 & MW & 81.046 & $-$24.525 & 12.9 & 18.8 & $-$1.60 & 13.0 & 10 & 39 & 16.0 & 0.03 & 5.4 & 4.6 & H & DeA05 & H \\
NGC 1978 &  & LMC & 82.188 & $-$66.236 & 50.1 & 50.0 & $-$0.38 & 1.9 & 18 & 40 & 18.4 & 0.21 & 5.1 & 4.6 & Mu08 & Mu07 & Fi92+Go06 \\
NGC 2004 &  & LMC & 82.678 & $-$67.286 & 50.1 & 50.0 & $-$0.58 & 0.02 & 6 & 22 & 16.2 & 0.36 & 4.3 & 3.9 & Ni15 & Ni15 & McL05 \\
NGC 2019 &  & LMC & 82.986 & $-$70.160 & 50.1 & 50.0 & $-$1.37 & 12.5 & 2 & 10 & 15.8 & 0.43 & 5.5 & 5.3 & Jo06 & Ol98 & McL05 \\
NGC 2100 &  & LMC & 85.538 & $-$69.212 & 50.1 & 50.0 & $-$0.42 & 0.02 & 4 & 18 & 15.6 & 0.65 & 4.4 & 4.1 & Pa16 & Ni15 & McL05 \\

\hline
\end{tabular}
\medskip
\emph{Notes}
Column (1): GC name.
Column (2): Other common identifiers for GC.
Column (3): Host galaxy.
Columns (4) and (5): Right ascension and declination in decimal degrees.
Values for MW star clusters are from the 2010 version of the \citet{1996AJ....112.1487H} catalogue; values for other galaxies are from NED.
Column (6): Heliocentric distance in kpc.
For the MW globular clusters, the distances are from the Harris catalogue.
For the Fornax dSph, the LMC and SMC star clusters we place all star clusters in each galaxy at the same distance, adopting the same distances as \citet{2005ApJS..161..304M}, namely 137 kpc, 50 kpc and 60 kpc respectively. 
Column (7): Galactocentric distance in kpc.
For MW globular clusters, the distances are from the Harris catalogue.
For the Fornax dSph, the LMC and SMC we adopt the same distances as \citet{2005ApJS..161..304M}, namely 149 kpc, 50 kpc and 61 kpc respectively.
Column (8): Metallicity in dex.
Column (9): Age in Gyr.
Column (10): Projected core radius in arcmin.
Column (11): Projected half-light radius in arcmin.
Column (12): $V$-band central surface brightness in mag per arcsec$^{-2}$.
Column (13): $V$-band extinction in mag.
Column (14): GC log stellar mass in solar masses calculated from $V$-band luminosity.
Column (15): Log stellar mass in solar masses enclosed by the WiFeS field-of-view calculated from the surface brightness profile.
Columns (16), (17) and (18): Sources for [Fe/H], age and structural parameters respectively. References are H: The 2010 version of the \citet{1996AJ....112.1487H} catalogue;
Ba09: \citet{2009A&A...507..405B};
Ba13: \citet{2013MNRAS.431L.122B};
Co11: \citet{2011ApJ...735...55C};
Da16: \citet{2016ApJ...829...77D};
Da98: \citet{1998AJ....115.1934D};
DeA05: \citet{2005AJ....130..116D};
Do10: \citet{2010ApJ...708..698D};
Do11: \citet{2011ApJ...738...74D};
Fi92+Go06: \citet{1992AJ....104.1086F} and \citet{2006MNRAS.369..697G};
Ge97: \citet{1997AJ....114.1920G};
Gl08a: \citet{2008AJ....135.1106G};
Gl08b: \citet{2008AJ....136.1703G};
Gl09: \citet{2009AJ....138.1403G};
Go06+09: \citet{2006MNRAS.369..697G} and \citet{2009AJ....137.4988G};
Go06+11: \citet{2006MNRAS.369..697G} and \citet{2011ApJ...737....3G};
Go14: \citet{2014ApJ...797...35G};
Gr06: \citet{2006AJ....132.1630G};
Hi99: \citet{1999A&A...345..430H};
Jo06: \citet{2006ApJ...640..801J};
Ke07: \citet{2007A&A...462..139K};
La12: \citet{2012A&A...546A..53L};
La14: \citet{2014ApJ...782...50L};
Ma03: \citet{2003MNRAS.338...85M};
McL05: \citet{2005ApJS..161..304M};
Me05: \citet{2006A&A...456.1085M};
Mi14: \citet{2014ApJ...785...21M};
Mi98: \citet{1998AJ....116.2395M};
Mu07: \citet{2007AJ....133.2053M};
Mu08: \citet{2008AJ....136..375M};
Mu10: \citet{2010ApJ...717..277M};
Mu11: \citet{2011MNRAS.413..837M};
Mu12: \citet{2012ApJ...746L..19M};
Ni15: \citet{2015A&A...575A..62N};
Ol91: \citet{1991AJ....101..515O};
Ol98: \citet{1998MNRAS.300..665O};
Or08: \citet{2008ApJ...687L..79O};
Pa16: \citet{2016MNRAS.458.3968P};
Si02: \citet{2002ApJ...579..275S};
Vi07: \citet{2007ApJ...663..296V};
Zo01: \citet{2001AJ....121.2638Z};
deB16: \citet{2016A&A...590A..35D};
\end{table}
\end{landscape}

\begin{landscape}
\begin{table}
\contcaption{General properties of the WAGGS sample}
\begin{tabular}{cccccccccccccccccc}
\hline
ID & Name & Galaxy & RA & Dec & $r_{\sun}$ & $r_{gc}$ & [Fe/H] & Age & $R_{c}$ & $R_{h}$ & $\mu_{V0}$ & $A_{V}$ & GC  & FoV  & [Fe/H] & Age & Structural \\
& & & & & & & & & & & & & Mass & Mass & Source & Source & Source \\
& & & [$^{\circ}$] & [$^{\circ}$] & [kpc] & [kpc] & [dex] & [Gyr] & [arcsec] & [arcsec] & [mag arcsec$^{-2}$] & [mag] & [$\log$ M$_{\sun}$]  & [$\log$ M$_{\sun}$] &  &  &   \\
(1) & (2) & (3) & (4) & (5) & (6) & (7) & (8) & (9) & (10) & (11) & (12) & (13) & (14) & (15) & (16) & (17) & (18) \\ \hline
NGC 2136 &  & LMC & 88.241 & $-$69.493 & 50.1 & 50.0 & $-$0.40 & 0.1 & 7 & 14 & 17.1 & 0.59 & 4.4 & 4.1 & Mu12 & Ni15 & McL05 \\
NGC 2808 &  & MW & 138.013 & $-$64.864 & 9.6 & 11.1 & $-$1.14 & 11.5 & 15 & 48 & 15.1 & 0.67 & 6.0 & 5.2 & H & Mi14 & H \\
NGC 3201 &  & MW & 154.403 & $-$46.412 & 4.9 & 8.8 & $-$1.59 & 12.0 & 78 & 186 & 19.0 & 0.73 & 5.2 & 3.3 & H & Do10 & H \\
NGC 4147 &  & MW & 182.526 & 18.543 & 19.3 & 21.4 & $-$1.80 & 12.8 & 5 & 29 & 17.4 & 0.06 & 4.7 & 4.2 & H & Do10 & H \\
NGC 4590 & M 68 & MW & 189.867 & $-$26.744 & 10.3 & 10.2 & $-$2.23 & 13.0 & 35 & 91 & 18.8 & 0.15 & 5.2 & 3.7 & H & Do10 & H \\
NGC 4833 &  & MW & 194.891 & $-$70.876 & 6.6 & 7.0 & $-$1.85 & 13.0 & 60 & 145 & 18.5 & 0.98 & 5.5 & 3.8 & H & Do10 & H \\
NGC 5024 & M 53 & MW & 198.230 & 18.168 & 17.9 & 18.4 & $-$2.10 & 13.2 & 21 & 79 & 17.4 & 0.06 & 5.7 & 4.7 & H & Do10 & H \\
NGC 5139 & $\omega$ Cen & MW & 201.697 & $-$47.480 & 5.2 & 6.4 & $-$1.53 & 11.0 & 142 & 300 & 16.8 & 0.37 & 6.3 & 4.1 & H & Vi07 & H \\
NGC 5272 & M 3 & MW & 205.548 & 28.377 & 10.2 & 12.0 & $-$1.50 & 12.5 & 22 & 139 & 16.6 & 0.03 & 5.8 & 4.5 & H & Do10 & H \\
NGC 5286 &  & MW & 206.612 & $-$51.374 & 11.7 & 8.9 & $-$1.69 & 13.0 & 17 & 44 & 16.2 & 0.73 & 5.7 & 5.0 & H & Do10 & H \\
NGC 5634 &  & MW & 217.405 & $-$5.976 & 25.2 & 21.2 & $-$1.88 & 13.0 & 5 & 52 & 17.2 & 0.15 & 5.3 & 4.5 & H & Me05 & H \\
NGC 5694 &  & MW & 219.901 & $-$26.539 & 35.0 & 29.4 & $-$1.98 & 13.6 & 4 & 24 & 16.5 & 0.27 & 5.4 & 4.9 & H & DeA05 & H \\
NGC 5824 &  & MW & 225.994 & $-$33.068 & 32.1 & 25.9 & $-$1.91 & 13.0 & 4 & 27 & 15.2 & 0.40 & 5.8 & 5.4 & H & DeA05 & H \\
NGC 5904 & M 5 & MW & 229.638 & 2.081 & 7.5 & 6.2 & $-$1.29 & 12.2 & 26 & 106 & 16.1 & 0.09 & 5.8 & 4.5 & H & Do10 & H \\
NGC 5927 &  & MW & 232.003 & $-$50.673 & 7.7 & 4.6 & $-$0.49 & 12.2 & 25 & 66 & 16.9 & 1.37 & 5.4 & 4.7 & H & Do10 & H \\
NGC 5986 &  & MW & 236.512 & $-$37.786 & 10.4 & 4.8 & $-$1.59 & 13.2 & 28 & 59 & 17.7 & 0.86 & 5.6 & 4.4 & H & Do10 & H \\
NGC 6093 & M 80 & MW & 244.260 & $-$22.976 & 10.0 & 3.8 & $-$1.75 & 13.5 & 9 & 37 & 15.1 & 0.55 & 5.5 & 5.0 & H & Do10 & H \\
NGC 6121 & M 4 & MW & 245.897 & $-$26.526 & 2.2 & 5.9 & $-$1.16 & 12.5 & 70 & 260 & 18.0 & 1.07 & 5.1 & 3.1 & H & Do10 & H \\
NGC 6139 &  & MW & 246.918 & $-$38.849 & 10.1 & 3.6 & $-$1.65 & -- & 9 & 51 & 17.2 & 2.29 & 5.6 & 4.8 & H & -- & H \\
NGC 6171 & M 107 & MW & 248.133 & $-$13.054 & 6.4 & 3.3 & $-$1.02 & 12.8 & 34 & 104 & 18.9 & 1.01 & 5.1 & 3.6 & H & Do10 & H \\
NGC 6218 & M 12 & MW & 251.809 & $-$1.949 & 4.8 & 4.5 & $-$1.37 & 13.2 & 47 & 106 & 18.1 & 0.58 & 5.2 & 3.5 & H & Do10 & H \\
NGC 6254 & M 10 & MW & 254.288 & $-$4.100 & 4.4 & 4.6 & $-$1.56 & 13.0 & 46 & 117 & 17.7 & 0.86 & 5.2 & 3.7 & H & Do10 & H \\
NGC 6266 & M 62 & MW & 255.303 & $-$30.114 & 6.8 & 1.7 & $-$1.18 & 12.5 & 13 & 55 & 15.1 & 1.44 & 5.9 & 5.1 & H & DeA05 & H \\
NGC 6273 & M 19 & MW & 255.657 & $-$26.268 & 8.8 & 1.7 & $-$1.74 & 13.2 & 26 & 79 & 16.8 & 1.16 & 5.9 & 4.8 & H & DeA05 & H \\
NGC 6284 &  & MW & 256.119 & $-$24.765 & 15.3 & 7.5 & $-$1.26 & 12.0 & 4 & 40 & 16.4 & 0.86 & 5.4 & 4.6 & H & DeA05 & H \\
NGC 6293 &  & MW & 257.542 & $-$26.582 & 9.5 & 1.9 & $-$1.99 & 13.0 & 3 & 53 & 15.7 & 1.10 & 5.3 & 4.4 & H & Me05 & H \\
NGC 6304 &  & MW & 258.634 & $-$29.462 & 5.9 & 2.3 & $-$0.45 & 12.8 & 13 & 85 & 17.2 & 1.65 & 5.1 & 4.3 & H & Do10 & H \\
NGC 6316 &  & MW & 259.155 & $-$28.140 & 10.4 & 2.6 & $-$0.45 & -- & 10 & 39 & 17.4 & 1.65 & 5.6 & 4.6 & H & -- & H \\
NGC 6333 & M 9 & MW & 259.797 & $-$18.516 & 7.9 & 1.7 & $-$1.77 & -- & 27 & 58 & 17.4 & 1.16 & 5.4 & 4.4 & H & -- & H \\
NGC 6342 &  & MW & 260.292 & $-$19.587 & 8.5 & 1.7 & $-$0.55 & 12.5 & 3 & 44 & 17.0 & 1.40 & 4.8 & 3.9 & H & DeA05 & H \\
NGC 6356 &  & MW & 260.896 & $-$17.813 & 15.1 & 7.5 & $-$0.40 & 12.8 & 14 & 49 & 17.0 & 0.86 & 5.6 & 4.9 & H & Me05 & H \\
NGC 6352 &  & MW & 261.371 & $-$48.422 & 5.6 & 3.3 & $-$0.64 & 13.0 & 50 & 123 & 18.2 & 0.67 & 4.8 & 3.7 & H & Do10 & H \\
NGC 6362 &  & MW & 262.979 & $-$67.048 & 7.6 & 5.1 & $-$0.99 & 12.5 & 68 & 123 & 19.3 & 0.27 & 5.0 & 3.3 & H & Do10 & H \\
NGC 6388 &  & MW & 264.072 & $-$44.735 & 9.9 & 3.1 & $-$0.55 & 11.8 & 7 & 31 & 14.5 & 1.13 & 6.0 & 5.3 & H & Mi14 & H \\
NGC 6397 &  & MW & 265.175 & $-$53.674 & 2.3 & 6.0 & $-$2.02 & 13.5 & 3 & 174 & 15.5 & 0.55 & 4.9 & 3.0 & H & Do10 & H \\
NGC 6440 &  & MW & 267.220 & $-$20.360 & 8.5 & 1.3 & $-$0.36 & -- & 8 & 29 & 17.1 & 3.27 & 5.7 & 5.1 & H & -- & H \\
NGC 6441 &  & MW & 267.554 & $-$37.051 & 11.6 & 3.9 & $-$0.46 & 12.0 & 8 & 34 & 14.9 & 1.44 & 6.1 & 5.5 & H & Mi14 & H \\
NGC 6522 &  & MW & 270.892 & $-$30.034 & 7.7 & 0.6 & $-$1.34 & 13.7 & 3 & 60 & 15.8 & 1.47 & 5.3 & 4.3 & H & Ba09 & H \\
NGC 6528 &  & MW & 271.207 & $-$30.056 & 7.9 & 0.6 & $-$0.11 & 11.0 & 8 & 23 & 16.7 & 1.65 & 4.9 & 4.5 & H & La14 & H \\
NGC 6541 &  & MW & 272.010 & $-$43.715 & 7.5 & 2.1 & $-$1.81 & 13.2 & 11 & 64 & 15.4 & 0.43 & 5.6 & 4.6 & H & Do10 & H \\
NGC 6553 &  & MW & 272.323 & $-$25.909 & 6.0 & 2.2 & $-$0.18 & 11.0 & 32 & 62 & 18.2 & 1.92 & 5.3 & 4.2 & H & Zo01 & H \\
NGC 6569 &  & MW & 273.412 & $-$31.827 & 10.9 & 3.1 & $-$0.76 & -- & 21 & 48 & 18.1 & 1.62 & 5.5 & 4.5 & H & -- & H \\
NGC 6584 &  & MW & 274.657 & $-$52.216 & 13.5 & 7.0 & $-$1.50 & 12.2 & 16 & 44 & 17.6 & 0.31 & 5.3 & 4.3 & H & Do10 & H \\
\hline
\end{tabular}
\end{table}
\end{landscape}

\begin{landscape}
\begin{table}
\contcaption{General properties of the WAGGS sample}
\begin{tabular}{cccccccccccccccccc}
\hline
ID & Name & Galaxy & RA & Dec & $r_{\sun}$ & $r_{gc}$ & [Fe/H] & Age & $R_{c}$ & $R_{h}$ & $\mu_{V0}$ & $A_{V}$ & GC  & FoV  & [Fe/H] & Age & Structural \\
& & & & & & & & & & & & & Mass & Mass & Source & Source & Source \\
& & & [$^{\circ}$] & [$^{\circ}$] & [kpc] & [kpc] & [dex] & [Gyr] & [arcsec] & [arcsec] & [mag arcsec$^{-2}$] & [mag] & [$\log$ M$_{\sun}$]  & [$\log$ M$_{\sun}$] &  &  &   \\
(1) & (2) & (3) & (4) & (5) & (6) & (7) & (8) & (9) & (10) & (11) & (12) & (13) & (14) & (15) & (16) & (17) & (18) \\ \hline
NGC 6624 &  & MW & 275.919 & $-$30.361 & 7.9 & 1.2 & $-$0.44 & 13.0 & 4 & 49 & 15.3 & 0.86 & 5.2 & 4.4 & H & Do10 & H \\
NGC 6637 & M 69 & MW & 277.846 & $-$32.348 & 8.8 & 1.7 & $-$0.64 & 12.5 & 20 & 50 & 16.8 & 0.55 & 5.3 & 4.4 & H & Do10 & H \\
NGC 6652 &  & MW & 278.940 & $-$32.991 & 10.0 & 2.7 & $-$0.81 & 13.2 & 6 & 29 & 16.1 & 0.27 & 4.9 & 4.3 & H & Do10 & H \\
NGC 6656 & M 22 & MW & 279.100 & $-$23.905 & 3.2 & 4.9 & $-$1.70 & 13.5 & 80 & 202 & 17.4 & 1.04 & 5.6 & 3.7 & H & Mi14 & H \\
NGC 6681 & M 70 & MW & 280.803 & $-$32.292 & 9.0 & 2.2 & $-$1.62 & 13.0 & 2 & 43 & 14.2 & 0.21 & 5.1 & 4.3 & H & Do10 & H \\
NGC 6715 & M 54 & MW & 283.764 & $-$30.480 & 26.5 & 18.9 & $-$1.49 & 13.2 & 5 & 49 & 14.8 & 0.46 & 6.2 & 5.7 & H & Mi14 & H \\
NGC 6717 & Pal 9 & MW & 283.775 & $-$22.701 & 7.1 & 2.4 & $-$1.26 & 13.0 & 5 & 41 & 16.8 & 0.67 & 4.5 & 3.8 & H & Do10 & H \\
NGC 6723 &  & MW & 284.888 & $-$36.632 & 8.7 & 2.6 & $-$1.10 & 12.8 & 50 & 92 & 18.1 & 0.15 & 5.4 & 3.9 & H & Do10 & H \\
NGC 6752 &  & MW & 287.717 & $-$59.985 & 4.0 & 5.2 & $-$1.54 & 12.5 & 10 & 115 & 14.9 & 0.12 & 5.3 & 4.2 & H & Do10 & H \\
NGC 6809 & M 55 & MW & 294.999 & $-$30.965 & 5.4 & 3.9 & $-$1.94 & 13.5 & 108 & 170 & 19.4 & 0.24 & 5.3 & 3.0 & H & Do10 & H \\
NGC 6838 & M 71 & MW & 298.444 & 18.779 & 4.0 & 6.7 & $-$0.78 & 12.5 & 38 & 100 & 19.8 & 0.76 & 4.5 & 2.8 & H & Do10 & H \\
NGC 6864 & M 75 & MW & 301.520 & $-$21.921 & 20.9 & 14.7 & $-$1.29 & 11.2 & 5 & 28 & 15.5 & 0.49 & 5.7 & 5.2 & H & Me05 & H \\
NGC 6934 &  & MW & 308.547 & 7.404 & 15.6 & 12.8 & $-$1.47 & 12.0 & 13 & 41 & 17.4 & 0.31 & 5.2 & 4.5 & H & Do10 & H \\
NGC 7006 &  & MW & 315.372 & 16.187 & 41.2 & 38.5 & $-$1.52 & 12.2 & 10 & 26 & 18.6 & 0.15 & 5.3 & 4.7 & H & Do11 & H \\
NGC 7078 & M 15 & MW & 322.493 & 12.167 & 10.4 & 10.4 & $-$2.37 & 13.2 & 8 & 60 & 14.2 & 0.31 & 5.9 & 5.3 & H & Do10 & H \\
NGC 7089 & M 2 & MW & 323.363 & $-$0.823 & 11.5 & 10.4 & $-$1.65 & 12.5 & 19 & 64 & 15.8 & 0.18 & 5.8 & 4.9 & H & Do10 & H \\
NGC 7099 & M 30 & MW & 325.092 & $-$23.180 & 8.1 & 7.1 & $-$2.27 & 13.2 & 4 & 62 & 15.4 & 0.09 & 5.2 & 4.1 & H & Do10 & H \\

\hline
\end{tabular}
\end{table}
\end{landscape}

\addtolength{\tabcolsep}{2pt}

\section{Observations \& Data Reduction}
\label{sec:observations}
We used the WiFeS integral field spectrograph \citep{2007Ap&SS.310..255D, 2010Ap&SS.327..245D} on the ANU 2.3 m telescope at the Siding Spring Observatory to observe the centres of 86 GCs in the MW and its satellite galaxies.
WiFeS uses an image slicer to reformat the telescope image into 25 slices, each 1 arcsec wide and 38 arcsec long on the detector.
This gives WiFeS a field-of-view of 38 by 25 arcsec which is similar to the median core radius of a MW GC \citep[20 arcsec, 2010 edition of ][]{1996AJ....112.1487H}.
The light from the image slicer is directed through a beam splitter to the red and blue arms of the spectrograph.
Each arm of the spectrograph uses a volume-phase holographic grating to disperse each slice as a long slit spectrum on a 4096 by 4096 pixel e2v CCD with 15 $\mu$m pixels.  
The plate scale in the spatial direction is 0.5 arcsec per pixel, so we binned every two spatial pixels to generate 1 by 1 arcsec spaxels.
We observed a single central pointing for each GC with two grating set-ups.
In one, we used the U7000 and R7000 gratings with the RT480 beam splitter to cover 3270 to 4350 \AA{} (0.27 \AA{} per pixel) and 5280 to 7020 \AA{} (0.44 \AA{} per pixel).
In the other, we used B7000 and I7000 gratings with the RT615 beam splitter to cover 4170 to 5540 \AA{} (0.37 \AA{} per pixel) and 6800 to 9050 \AA{} (0.57 \AA{} per pixel).
All four gratings give spectral resolutions of $\delta \lambda / \lambda \sim 6800$ and slightly undersample the line spread function.
This corresponds to a velocity dispersion of 19 km s$^{-1}$ which is similar to the velocity dispersions of the most massive GCs in the MW and M31 \citep[e.g.][]{1996AJ....112.1487H, 2011AJ....142....8S}.
The improvement in spectral resolution over previous studies is illustrated in Figure \ref{fig:effect_resolution}.

Our observations span 19 nights from January 2015 to October 2016.
Details of the observations are given in the observation log in Table \ref{tab:observing_log}, the full version of which is available online as part of the supplementary material.
The seeing generally ranged between 1.6 and 2.0 arcsec and conditions were often not photometric.

We used WiFeS in nod-and-shuffle \citep{2001PASP..113..197G} mode to perform accurate sky subtraction.
We observed cycles of 30 s on target and 30 s on sky with the total exposure times estimated from surface brightness profiles calculated from the structural parameters given in Table \ref{tab:sample_properties} using the \textsc{limepy} code \citep{2015MNRAS.454..576G} and a \citet{1966AJ.....71...64K} profile.
Blank patches of sky $\sim 10$ arcmin (on average 8 half-light radii) from the GC centres were used as the sky fields.
Coordinates of the object and sky fields are given in the observing log.
We note that we must not only accurately subtract the sky background but also the starlight and nebulosity of the host galaxy at the same location as the GC.
For GCs in the MW halo this is not a major issue, while for GCs in the bulge and near the disc of the MW , as well as near the LMC and SMC centres, it can be relevant.

Our usual observing strategy was to take three science exposures with the U7000 and R7000 gratings before exposing a NeAr arc with the same grating set-up.
We then switched the gratings to B7000 and I7000 and exposed a NeAr arc with the new grating set-up before taking three science exposures.
We then moved to the next target, took three B7000/I7000 science exposures and a B7000/I7000 arc before reconfiguring the gratings and observing the U7000/R7000 arc and the three U7000/R7000 science exposures.
Before each observing night we observed 5 bias, lamp flat and wire exposures (to spatially align the instrument) for each of the two grating set-ups.
We also observed sky flat exposures for both grating set-ups during twilight.

We typically observed a white dwarf and a metal poor red giant each night as spectrophotometric standards.
During our April 2016 and September/October 2016 runs, we observed a number of stars from both the MILES spectral library \citep{2006MNRAS.371..703S} and the Lick index standard stars \citep{1994ApJS...94..687W} to test the reliability of our spectral index measurements.
We did not use nod-and-shuffle for either the spectrophotometric standard star observations or the line index standards.
Details of the standard star observations are given in the observing log in Table \ref{tab:observing_log}.

\begin{table*}
\caption{Observing Log}
\begin{tabular}{cccccccc}
\hline
ID & Object & Sky & Date & Grating & Exposure & S/N & Notes \\
& & & & & [s] & [\AA${^{-1}}$] & \\
(1) & (2) & (3) & (4) & (5) & (6) & (7) & (8) \\ \hline
NGC 104 & 00:24:08.2 -72:04:52.1 & 00:08:51.3 -72:01:55 & 2015-01-30 10:50:09.5 & R7000 & 240 & 1145 & -- \\
NGC 104 & 00:24:08.2 -72:04:52.1 & 00:08:51.3 -72:01:55 & 2015-01-30 10:50:09.5 & U7000 & 240 & 264 & -- \\
NGC 104 & 00:24:08.2 -72:04:52.1 & 00:08:51.3 -72:01:55 & 2015-01-30 11:13:50.5 & B7000 & 180 & 689 & -- \\
NGC 104 & 00:24:08.2 -72:04:52.1 & 00:08:51.3 -72:01:55 & 2015-01-30 11:13:50.5 & I7000 & 180 & 911 & -- \\
NGC 362 & 01:03:16.9 -70:51:04.6 & 01:01:38.1 -70:42:12 & 2015-01-30 11:40:33.5 & B7000 & 180 & 338 & -- \\
NGC 362 & 01:03:16.9 -70:51:04.6 & 01:01:38.1 -70:42:12 & 2015-01-30 11:40:33.5 & I7000 & 180 & 470 & -- \\
NGC 362 & 01:03:16.9 -70:51:04.6 & 01:01:38.1 -70:42:12 & 2015-01-30 12:17:27.5 & U7000 & 180 & 91 & -- \\
NGC 362 & 01:03:16.9 -70:51:04.6 & 01:01:38.1 -70:42:12 & 2015-01-30 12:17:27.5 & R7000 & 180 & 487 & -- \\
HD 44007 & 06:18:48.2 -14:50:42 & 06:18:48.2 -14:50:42 & 2015-01-30 12:35:09.5 & U7000 & 210 & 366 & flux standard \\
HD 44007 & 06:18:48.2 -14:50:42 & 06:18:48.2 -14:50:42 & 2015-01-30 12:35:09.5 & R7000 & 210 & 904 & flux standard \\
HD 44007 & 06:18:48.2 -14:50:42 & 06:18:48.2 -14:50:42 & 2015-01-30 12:59:20.5 & I7000 & 180 & 774 & flux standard \\
HD 44007 & 06:18:48.2 -14:50:42 & 06:18:48.2 -14:50:42 & 2015-01-30 12:59:20.5 & B7000 & 180 & 732 & flux standard \\
... & ... & ... & ... & ... & ... & ... & ... \\ \hline
\end{tabular}

\medskip
\emph{Notes}
The full version of this table is provided in a machine readable form in the online Supporting Information.
Column (1): Globular cluster or star.
Column (2): Object coordinate.
Column (3): Sky coordinate.
Column (4): Observation date and UTC time.
Column (5): Grating.
Column (6): Exposure time in seconds.
Column (7): Mean signal-to-noise ratio per \AA{} for the integrated spectra. For the U7000 grating this was measured between 4000 and 4050 \AA{}; for the B7000 grating between 4800 and 4850 \AA{}; for the R7000 grating between 6400 and 6450; for the I7000 grating between 8400 and 8450 \AA{}.
\label{tab:observing_log}
\end{table*}

\subsection{Data reduction}
We used the \textsc{PyWiFeS} \citep{2014ascl.soft02034C, 2014Ap&SS.349..617C} pipeline to reduce the observations.
We provide an outline of the data reduction procedure performed by \textsc{PyWiFeS} here but refer the interested reader to \citet{2014Ap&SS.349..617C} for more details.
After performing bias subtraction, the pipeline separates each image into the 25 slitlets with nod-and-shuffle exposures being further divided into sky and object slitlets.
For each slitlet the pipeline calculates spectral flat field responses using the lamp flats, spatial flat field responses using the sky flats and wavelength solutions using the arc frames.
For each slitlet the pipeline performs cosmic ray identification and repair using a modified version of LA Cosmic \citep{2001PASP..113.1420V}.
For exposures using nod-and-shuffle, each sky slitlet is subtracted from the corresponding object slitlet.
Multiple exposures of the same object are combined slitlet by slitlet.
After applying the flat field and wavelength calibrations, the pipeline uses the wire exposures to spatially align the slitlets and corrects for the effects of atmospheric differential refraction using the equations of \citet{1982PASP...94..715F}.
The pipeline then resamples and combines the observed pixels of each slitlet in to a single rectilinear datacube.
The pipeline uses the observed standard stars to flux calibrate the datacubes.
When no standard star was observed, we used the star observed on the previous night.
For the R7000 and I7000 gratings, the pipeline fits the standard star spectra outside of wavelength regions affected by telluric absorption with a polynomial before using the ratio of the fitted polynomial to the observed standard star spectra to correct each datacube for telluric absorption.
We preferred to use the white dwarf standard stars for the telluric corrections, only using the red giant standard stars when no telluric star was observed that night.

We performed astrometry of the datacubes by comparing images created by summing the I7000 cubes along the wavelength direction with J-band 2MASS \citep{2006AJ....131.1163S} images. 
The I7000 cubes were used due to the stronger surface brightness fluctuations signal in the red while 2MASS imaging has similar spatial resolution to the WiFeS datacubes.
Accurate astrometry is, however, challenging for these targets, as the field is severely crowded at the effective spatial resolution and field size of the WiFeS IFU.

\subsection{Integrated Spectra}

For most of the GCs in our sample, the WiFeS field-of-view is smaller than their half-light diameter.
For these objects, we created integrated spectra by simply summing the spatial pixels, excluding the first two and last two rows of cubes as these are noisier (these rows come from the ends of each slice) and show larger residuals from the subtraction of sky emission lines. 
For the small number of GCs wth half-light diameters smaller than the instrument field-of-view (namely the 3 GCs in Fornax, NGC 416 in the SMC, NGC 1786, NGC 1868, NGC 1916, NGC 2004 and NGC 2136 in the LMC), we created integrated spectra by summing all the spatial pixels within their half-light radii.

\begin{figure}
\begin{center}
\includegraphics[width=240pt]{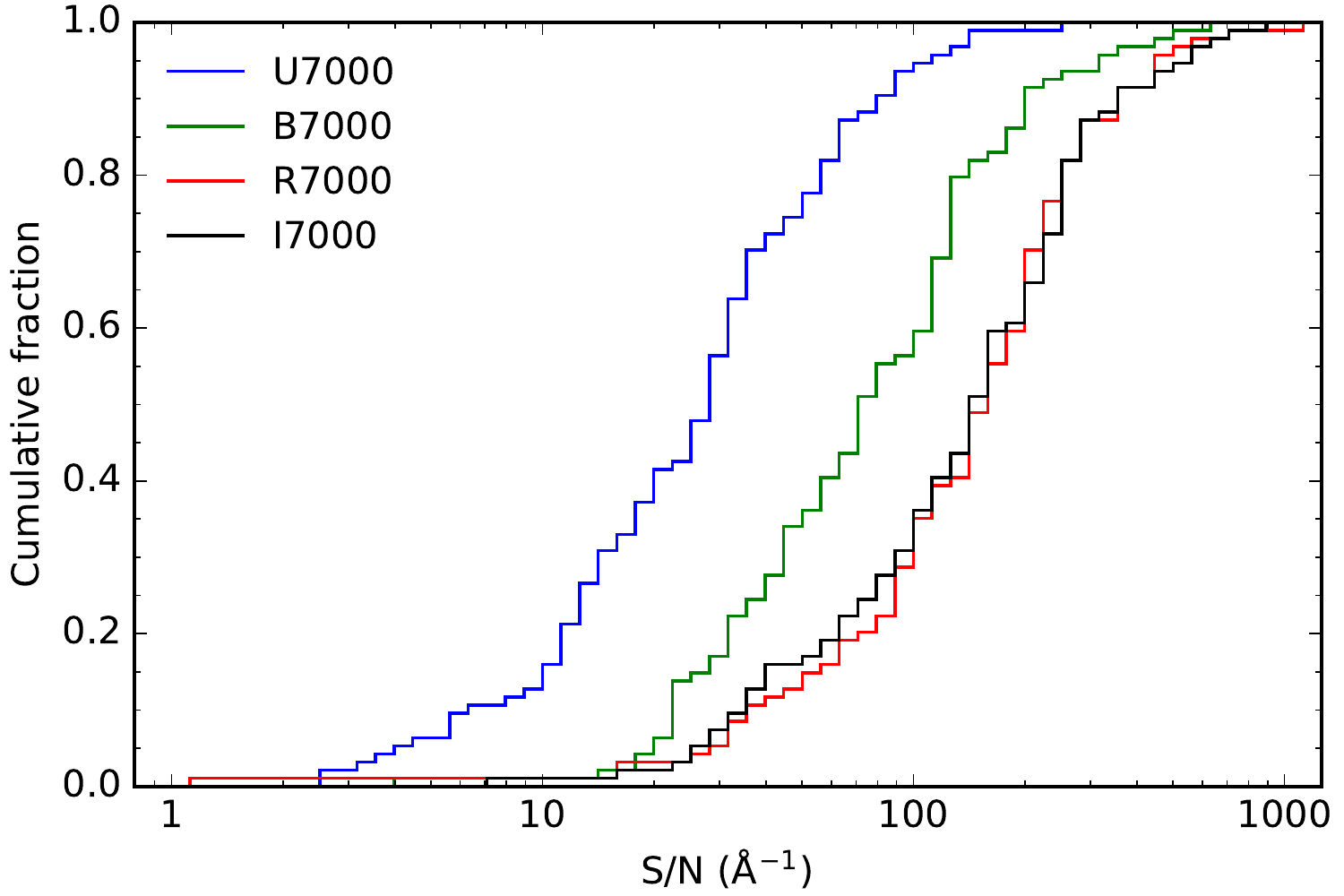}
\caption{Cumulative S/N per \AA{} distributions of the integrated spectra for each of the gratings.
In general the S/N of the U7000 spectra is lower than the B7000 spectra while the R7000 and I7000 are higher.}
\label{fig:s2n}
\end{center}
\end{figure}

For each integrated spectrum, we calculated the mean S/N in a grating- dependent wavelength region.
The wavelength regions, minimum, median and maximum S/Ns are given for each grating in Table \ref{tab:s2n} while the S/N distributions are plotted in Figure \ref{fig:s2n}.
The S/Ns for each observation are given in Table \ref{tab:observing_log}.
Generally, the U7000 grating has the worst S/N and the R7000 and I7000 gratings the best.
Since exposure times were calculated using $V$-band surface brightnesses, metal rich GCs and GCs with more foreground extinction generally have lower S/Ns at bluer wavelengths.
Except for the U7000 grating and wavelength regions affected by strong sky emission lines or telluric absorption, S/N generally does not vary dramatically within a grating.
In the case of the U7000 grating, the S/N increases significantly at shorter wavelengths.
The S/N of spectra of GCs in the MW satellites are generally lower than those in the MW with the median B7000 S/N being 115 \AA${^{-1}}$ in the MW and 46 \AA${^{-1}}$ in the MW satellites.

\begin{figure*}
\begin{center}
\includegraphics[width=504pt]{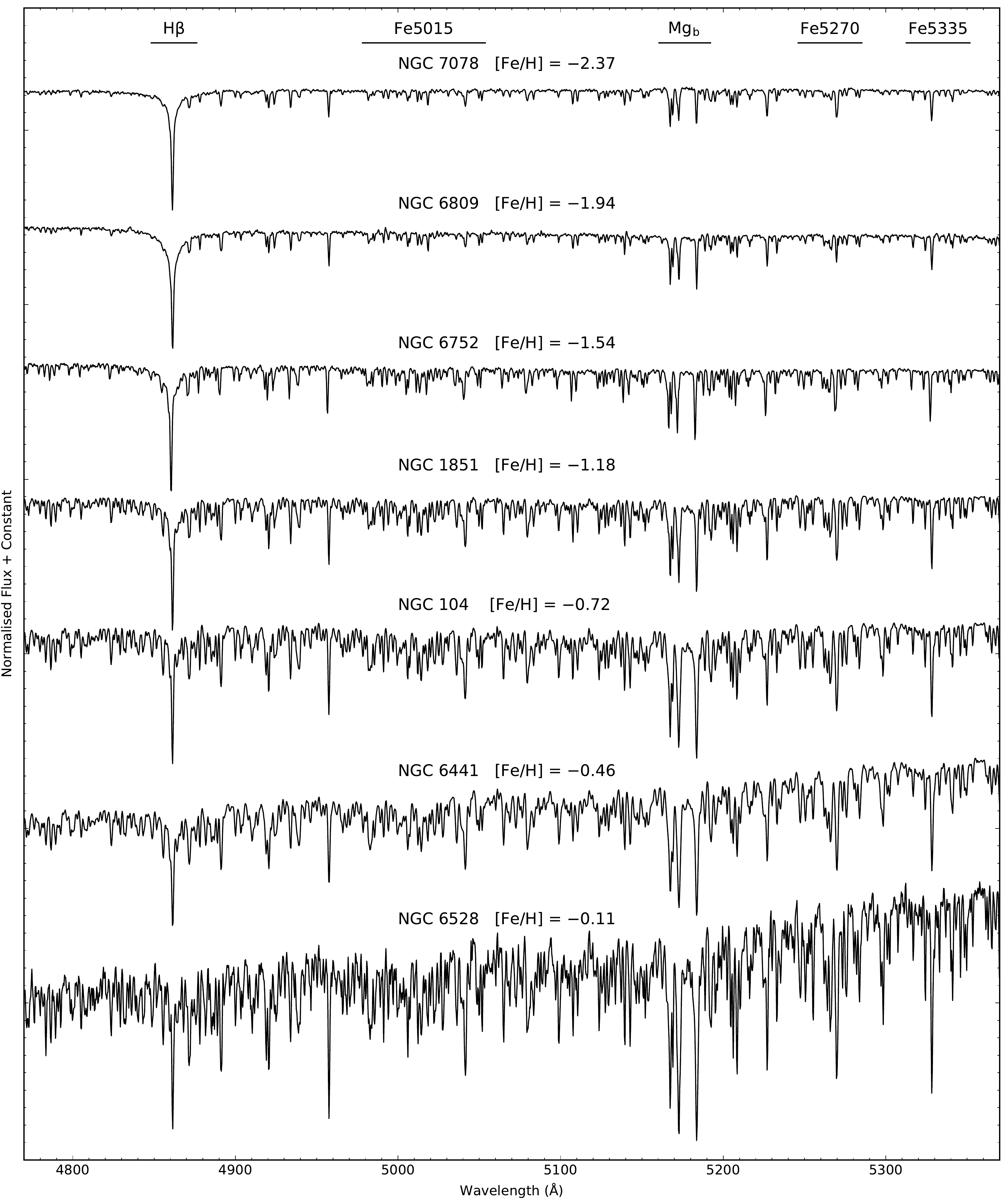}
\caption{WAGGS spectra in the region of H$\beta$ and Mg$_{b}$ for GCs with a range of metallicities.
The spectra increase in metallicity from top to bottom and have been offset an arbitrary amount in the y-axis.
The GCs plotted in this figure span the range of metallicities in the WAGGS sample and are all old (age $> 10$ Gyr).
A whole host of metal lines increase in strength with metallicity while H$\beta$ deceases.}
\label{fig:blue_metal}
\end{center}
\end{figure*}

\begin{figure*}
\begin{center}
\includegraphics[width=504pt]{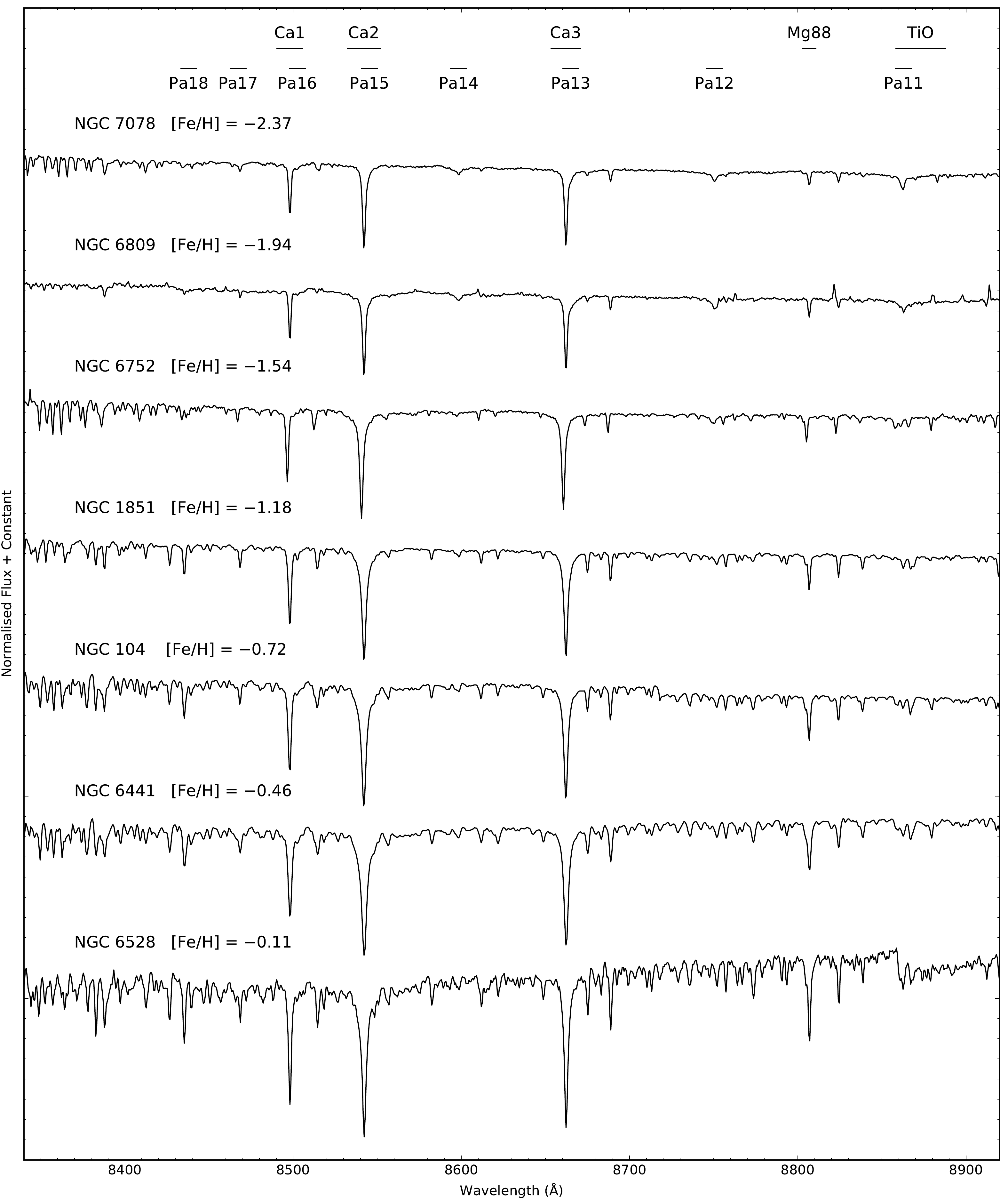}
\caption{WAGGS spectra in the region of the CaT for GCs with a range of metallicities.
The spectra increase in metallicity from top to bottom and have been offset an arbitrary amount in the y-axis.
The GCs plotted in this figure span the range of metallicities in the WAGGS sample and are all old (age $> 10$ Gyr).
Weak Paschen line absorption is only visible in the lowest metallicity GCs while the TiO bandhead at 8860 \AA{} only appears at high metallicity.
}
\label{fig:red_metal}
\end{center}
\end{figure*}

\begin{figure*}
\begin{center}
\includegraphics[width=504pt]{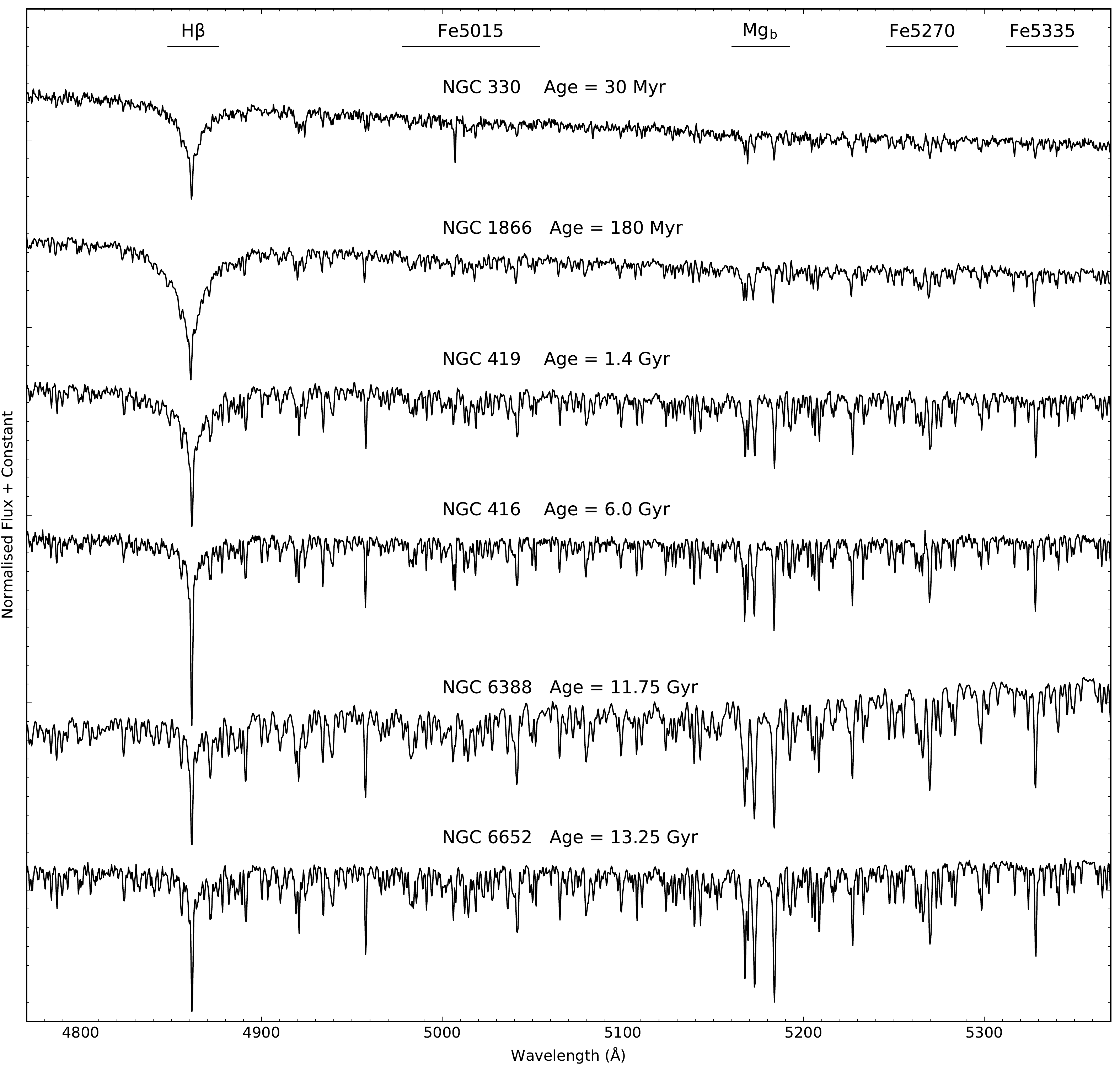}
\caption{WAGGS spectra in the region of H$\beta$ and Mg$_{b}$ for GCs with a range of ages.
The spectra increase in age from top to bottom and have been offset an arbitrary amount in the y-axis.
The GCs plotted in this figure span the range of ages in the WAGGS sample and have been selected to fall in the metallicity range $-1 <$ [Fe/H] $< -0.4$.
Going from young ages to old, the strength of H$\beta$ decreases and the strength of metal lines increases.}
\label{fig:blue_age}
\end{center}
\end{figure*}

\begin{figure*}
\begin{center}
\includegraphics[width=504pt]{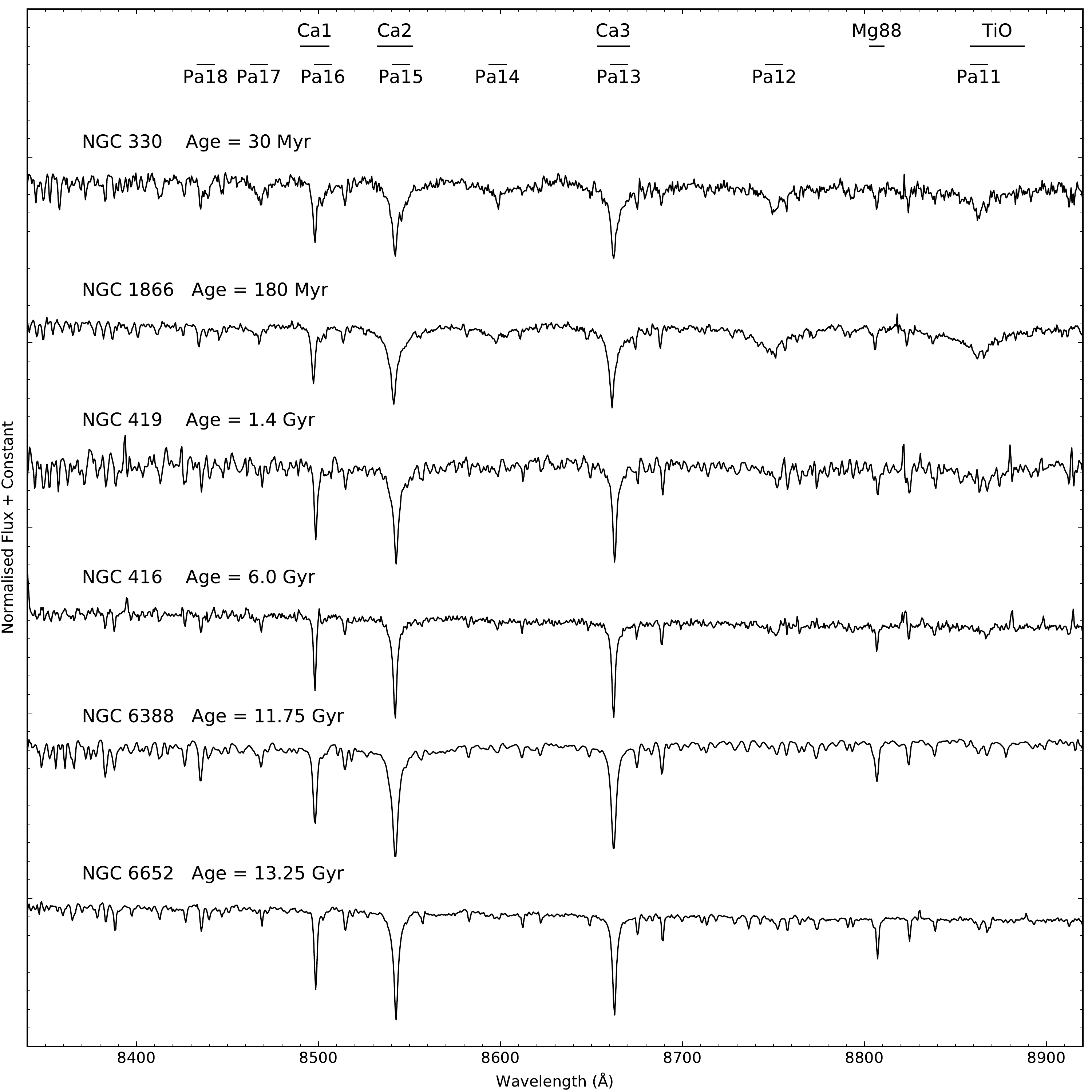}
\caption{WAGGS spectra in the region of the CaT for GCs with a range of ages.
The spectra increase in age from top to bottom and have been offset an arbitrary amount in the y-axis.
The GCs plotted in this figure span the range of ages in the WAGGS sample and have been selected to fall in the metallicity range $-1 <$ [Fe/H] $< -0.4$.
Note the presence of Paschen absorption only in the youngest GCs.}
\label{fig:red_age}
\end{center}
\end{figure*}

The integrated spectra of NGC 104 across the entire observed wavelength range are shown in Figure \ref{fig:ngc104}.
This illustrates the wide wavelength coverage of the WAGGS spectra which cover a wide range of spectral features including molecular bands in the near-UV, the traditional Lick \citep{1994ApJS...94..687W} indices, the sodium doublets at 5895 and 8910 \AA{}, H$\alpha$ and the CaT.
Integrated spectra of GCs with a range of metallicities in the spectral region of H$\beta$ and Mg$_{b}$ are shown in Figure \ref{fig:blue_metal} and around the CaT in Figure \ref{fig:red_metal}.
Unsurprisingly, in both figures a range of metal lines increase in strength with increasing metallicity.
The shape of the pseudo-continuum also becomes redder with increasing metallicity although we note that we have not corrected our spectra for the effects of extinction and that our more metal rich GCs generally have higher foreground reddening than our metal poor GCs.
The strength of H$\beta$ decreases with increasing metallicity where as in the CaT region, weak Paschen absorption is only seen at the lowest metallicities.
Integrated spectra of GCs with a range of ages in the spectral region of H$\beta$ and Mg$_{b}$ are shown in Figure \ref{fig:blue_age} and around the CaT in Figure \ref{fig:red_age}.
As expected, the strength H$\beta$ decreases with increasing age while the strength of metal lines increase with age.
In the region of the CaT, strong Paschen absorption is only seen in GCs younger than 1 Gyr.

\begin{figure*}
\begin{center}
\includegraphics[width=504pt]{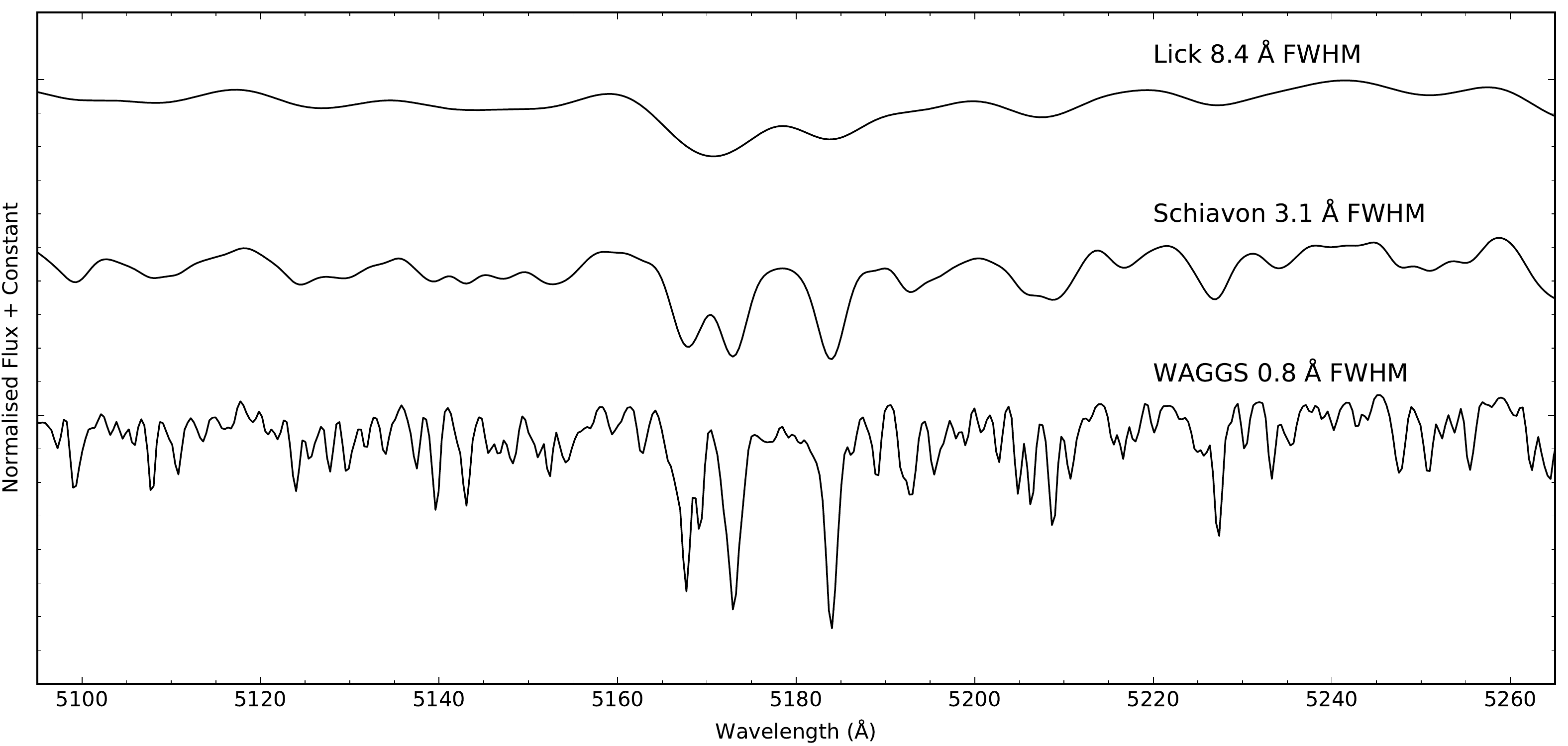}
\caption{WAGGS spectral resolution compared to previous work.
Our spectrum of NGC 104 around Mg$_{b}$ is plotted at its original spectral resolution ($R = 6800$, 0.8 \AA{} FWHM, bottom), smoothed to the spectral resolution of \citet{2005ApJS..160..163S} library of GC spectra ($R = 1700$, 3.1 \AA{} FWHM, middle) and of the Lick index system \citep[$R = 600$, 8.4 \AA{} FWHM, top]{1994ApJS...94..687W}.
The smoothed spectra have been offset in flux for legibility.
This NGC 104 spectrum has a S/N of 689 \AA$^{-1}$.
Our spectra have four times higher spectral resolution compared to the spectra of \citet{2005ApJS..160..163S} and eleven times the spectral resolution of the Lick system.
This high resolution allows us to study much weaker spectral lines.
Note that the apparently smooth pseudo-continuum at low resolution is made up of numerous weak absorption lines.}
\label{fig:effect_resolution}
\end{center}
\end{figure*}

\begin{table}
\caption{Signal-to-noise ratios}
\begin{tabular}{cccccc}
\hline
Grating & $\lambda_{min}$ & $\lambda_{max}$ & S/N$_{min}$ & S/N$_{med}$ & S/N$_{max}$ \\
& \AA{} & \AA{} & \AA${^{-1}}$ & \AA${^{-1}}$ & \AA${^{-1}}$ \\
(1) & (2) & (3) & (4) & (5) & (6) \\ \hline
U7000 & 4000 & 4050 & 0.3 & 29 & 264 \\ 
B7000 & 4800 & 4850 & 4.4 & 77 & 689 \\
R7000 & 6400 & 6450 & 1.2 & 161 & 1145 \\
I7000 & 8400 & 8450 & 7.9 & 157 & 911 \\ \hline
\end{tabular}

\medskip
\emph{Notes}
Column (1): Grating.
Column (2): Minimum wavelength of S/N calculation region.
Column (3): Maximum wavelength of S/N calculation region.
Column (4): Minimum S/N per \AA.
Column (5): Median S/N per \AA.
Column (6): Maximum S/N per \AA.
\label{tab:s2n}

\end{table}

\begin{figure*}
\begin{center}
\includegraphics[width=504pt]{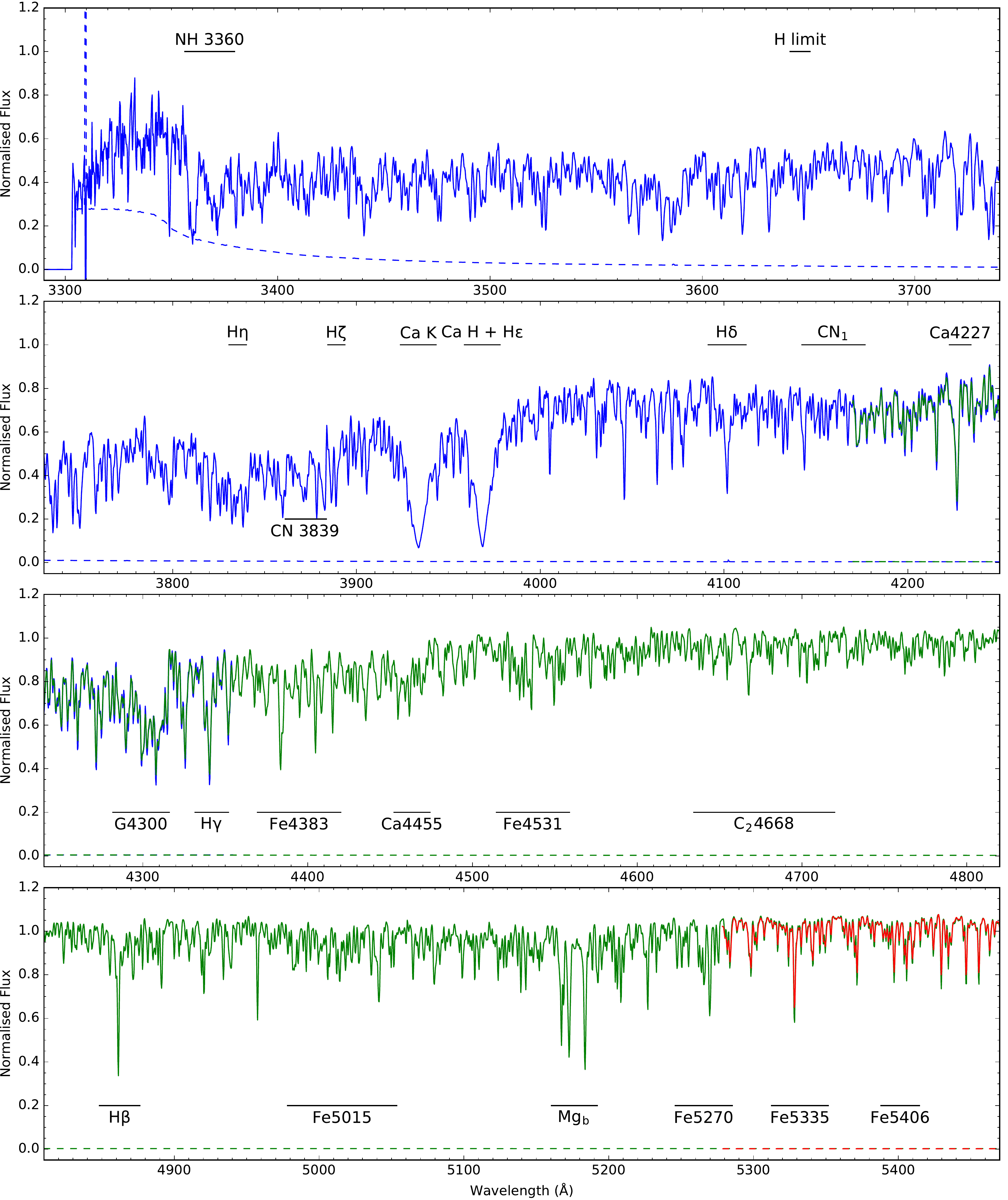}
\caption{WAGGS spectra of NGC 104.
The U7000 grating spectrum is shown in blue, the B7000 spectrum in green, the R7000 in red and the I7000 in black.
For each grating, the uncertainties provided by the \textsc{PyWiFeS} pipeline is shown as a dashed line.
Common used spectral indices are shown as black horizontal lines.
Grey shaded regions denote wavelength regions were telluric lines have been corrected by the \textsc{PyWiFeS} pipeline.
The spectra have been shifted to the rest frame and normalised such that the average flux in the wavelength range 5300 \AA{} to 5500 \AA{} is unity and that the average flux in the regions of overlap is the same for both gratings.}
\label{fig:ngc104}
\end{center}
\end{figure*}

\begin{figure*}
\begin{center}
\includegraphics[width=504pt]{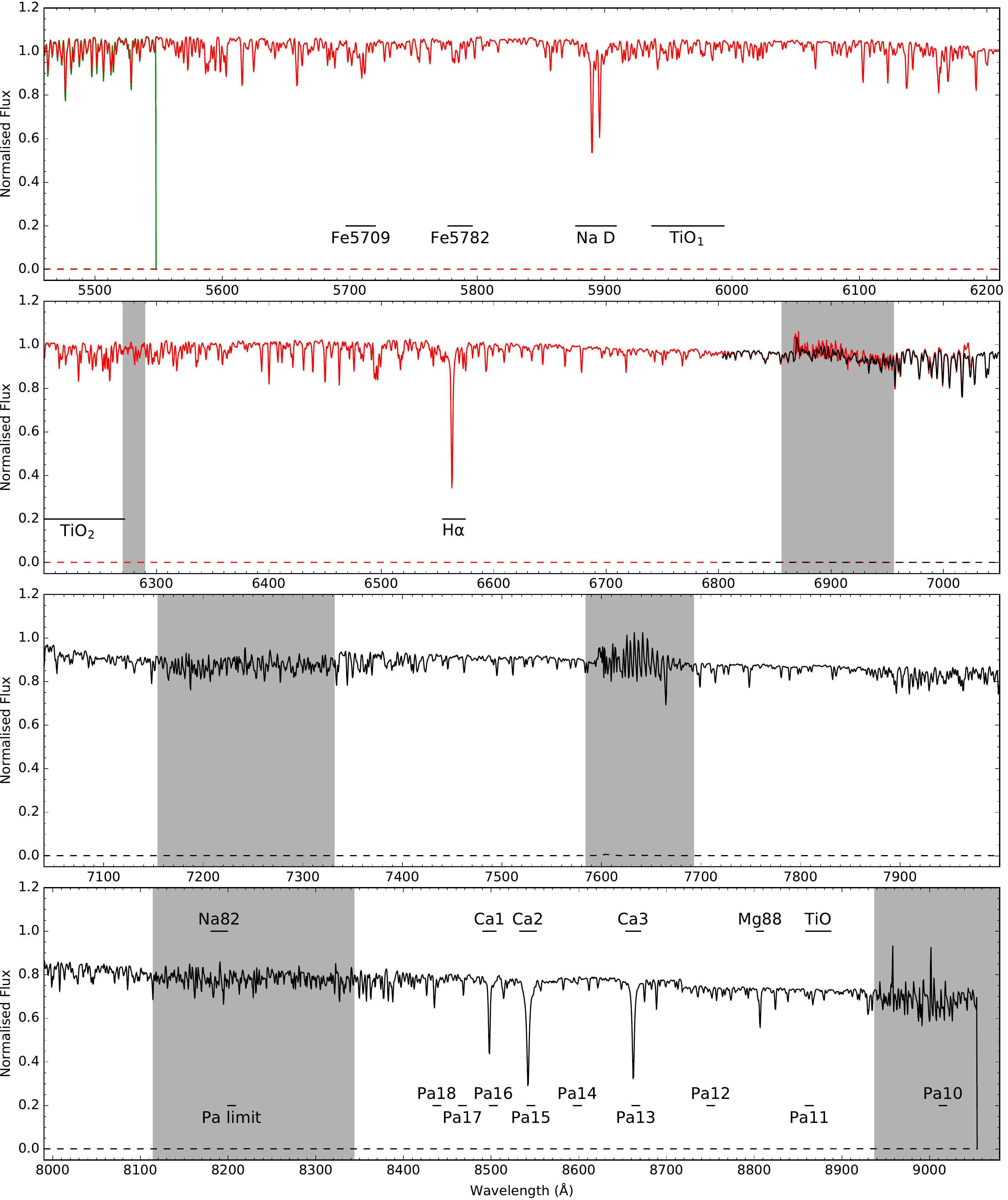}
\contcaption{WAGGS spectra of NGC 104.}
\end{center}
\end{figure*}

\subsection{Repeated observations}
In several cases we observed the same GC more than once with the same grating set-up.
Often this occurred when we were only able to observe one grating set-up for a GC on a given night; when revisiting those GCs in order to complete  the observations we usually elected to observe both grating set-ups.
In some cases, poor data quality prompted us to reobserve a particular GC.
We intentionally reobserved NGC 2808 and NGC 7099 to assess the repeatability of our observations.
We have repeated observations of the U7000 and R7000 gratings for NGC 2808, NGC 3201, NGC 5286, NGC 6304, NGC 6723 and NGC 7099 and of the B7000 and I7000 gratings for NGC 1846, NGC 2808, NGC 6284, NGC 6304, NGC 6342, NGC 6717 and NGC 7099.

\begin{figure*}
\begin{center}
\includegraphics[width=504pt]{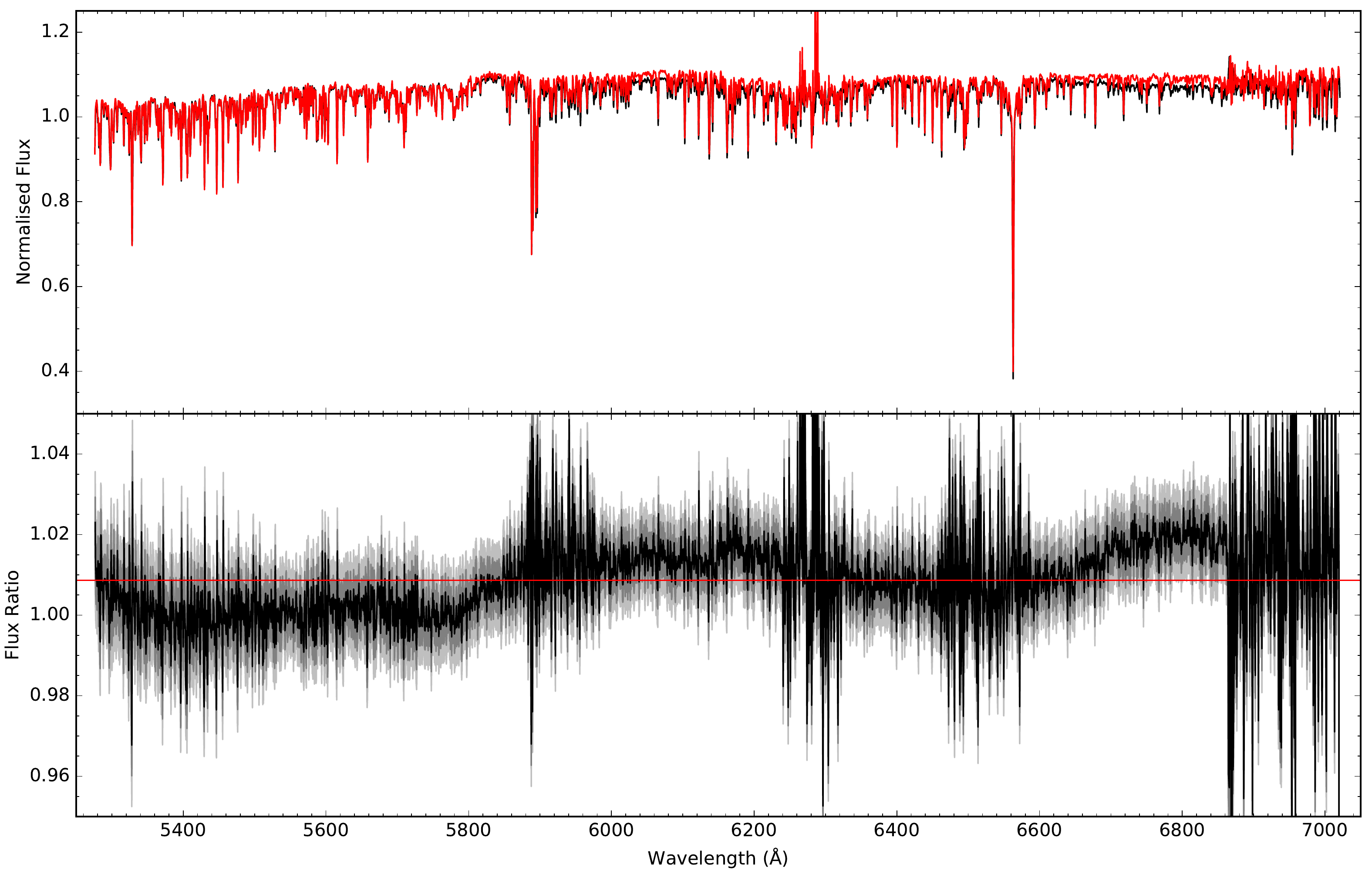}
\caption{\emph{Top:} Comparison of the R7000 integrated spectra of NGC 2808 from 2015-01-31 (black) and from 2015-07-09 (red).
This a relative comparison as both observations have been normalise such that their mean flux between 5300 \AA{} to 5500 \AA{} is unity.
The 2015-01-31 spectrum has a S/N of 744 \AA$^{-1}$ while the 2015-07-09 spectrum has a S/N of 389 \AA$^{-1}$.
These S/N are based on the uncertainties provided by the \textsc{PyWiFeS} pipeline.
The spectra have been shifted to the rest frame.
\emph{Bottom:} Ratio of the two nights observations in black.
The dark grey and light grey regions show the one and two sigma uncertainties in the ratio.
We note that the regions of greatest variance - 5880 to 5980 \AA{}, 6270 to 6330 \AA{}, 6460 to 6580 \AA{} and redwards of 6860 \AA{} - lie in wavelength regions affected by telluric lines.}
\label{fig:repeat_wide}
\end{center}
\end{figure*}

\begin{figure}
\begin{center}
\includegraphics[width=240pt]{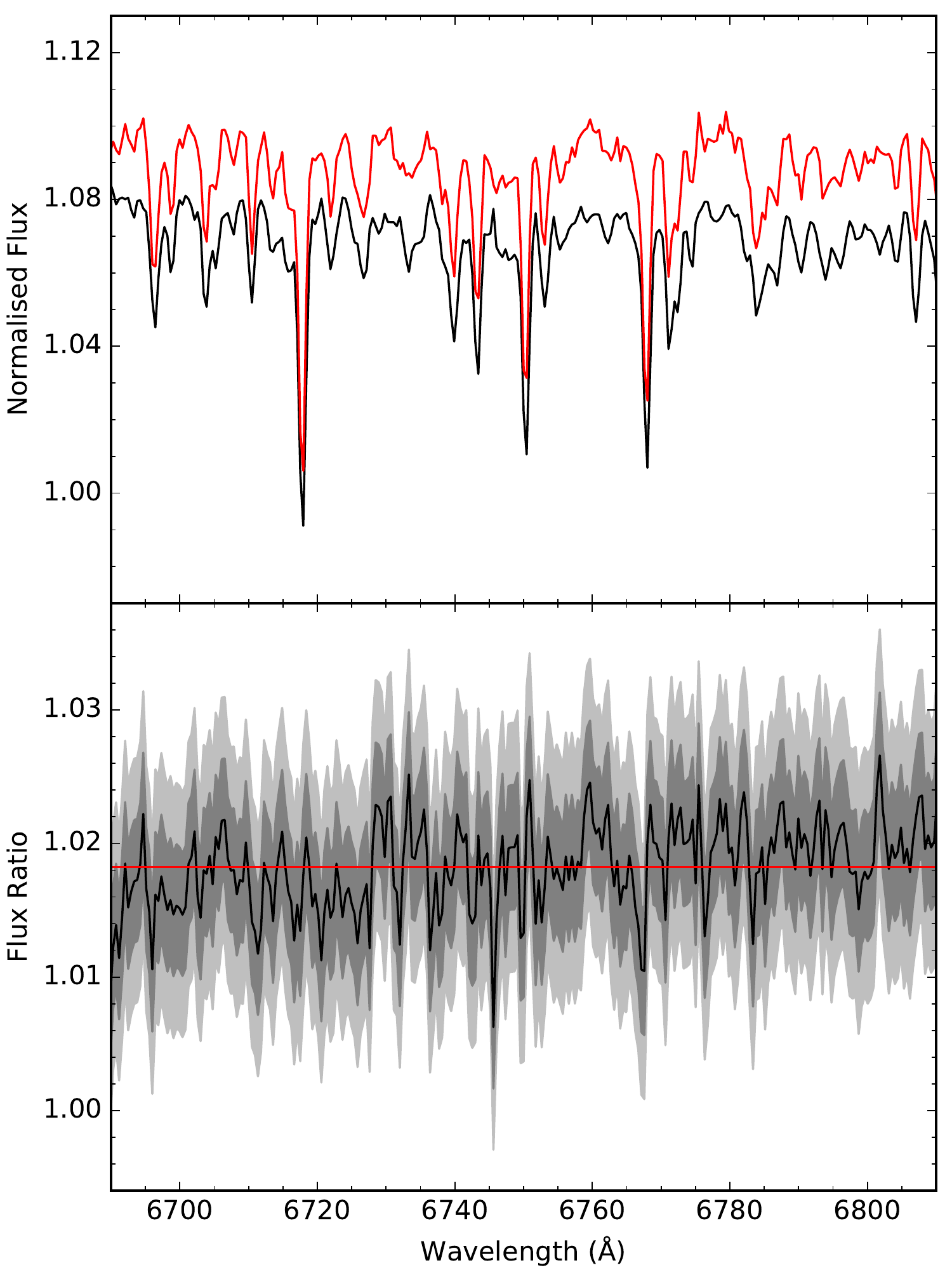}
\caption{\emph{Top:} Comparison of the R7000 spectra of NGC 2808 from 2015-01-31 (black) and from 2015-07-09 (red) over a narrower wavelength range than Figure \ref{fig:repeat_wide}.
As before the spectra have been normalised to unity in the wavelength range of 5300 \AA{} to 5500 \AA{} and shifted to the rest frame.
\emph{Bottom:} Ratio of the two nights observations in black.
The dark grey and light grey regions show the one and two sigma uncertainties in the ratio.
Modulo a $\sim 1$ per cent difference in flux calibration, the spectra from the two nights shows excellent agreement within the uncertainties provided by the \textsc{PyWiFeS} pipeline.}
\label{fig:repeat_narrow}
\end{center}
\end{figure}

We show a comparison of our NGC 2808 integrated spectra for two different nights (2015-01-31 and 2015-07-09) across the entire R7000 grating in Figure \ref{fig:repeat_wide} and over a narrower wavelength range in Figure \ref{fig:repeat_narrow}.
Since the declinations of the two observations differ by 3 arcsec, for this comparison we extracted spectra from spaxels which cover the same area on sky.
Over the entire grating, the flux calibration is generally consistent between observations to within a couple percent.
The largest variance in the ratio of the two nights' spectra is seen over wavelength regions such as redwards of 6875 \AA{} and around 6280 \AA , which are affected by telluric absorption.
For wavelength ranges of $\sim 100$ \AA{} or less, the differences between observations are generally consistent with the uncertainties provided by the \textsc{PyWiFeS} pipeline modulo a multiplicative factor in flux.
We see similar differences with the other gratings and other GCs, although we note that the quality of the flux calibration is poorer with the U7000 grating.
We believe that the \textsc{PyWiFeS} flux calibration procedure is responsible most of the systematic differences between repeated observations.
We aim to improve upon the flux calibration in future work.

We also note that due to effects of variable stars, in particular long period variables, the integrated light of globular clusters is intrinsically variable at the percent level on the timescale of hours to years \citep{2015Natur.527..488C}.
This effect is most noticeable in spectral features such as the TiO bands which are predominantly formed in the coolest, brightest giants.
We were able to trace a one percent difference between our two R7000 observations of NGC 2808 (observed 159 days apart) to three known \citep{2011A&A...529A.137L} long period variables in our field-of-view.

\subsection{Resolved stars and stochastic sampling}
\begin{figure}
\begin{center}
\includegraphics[width=240pt]{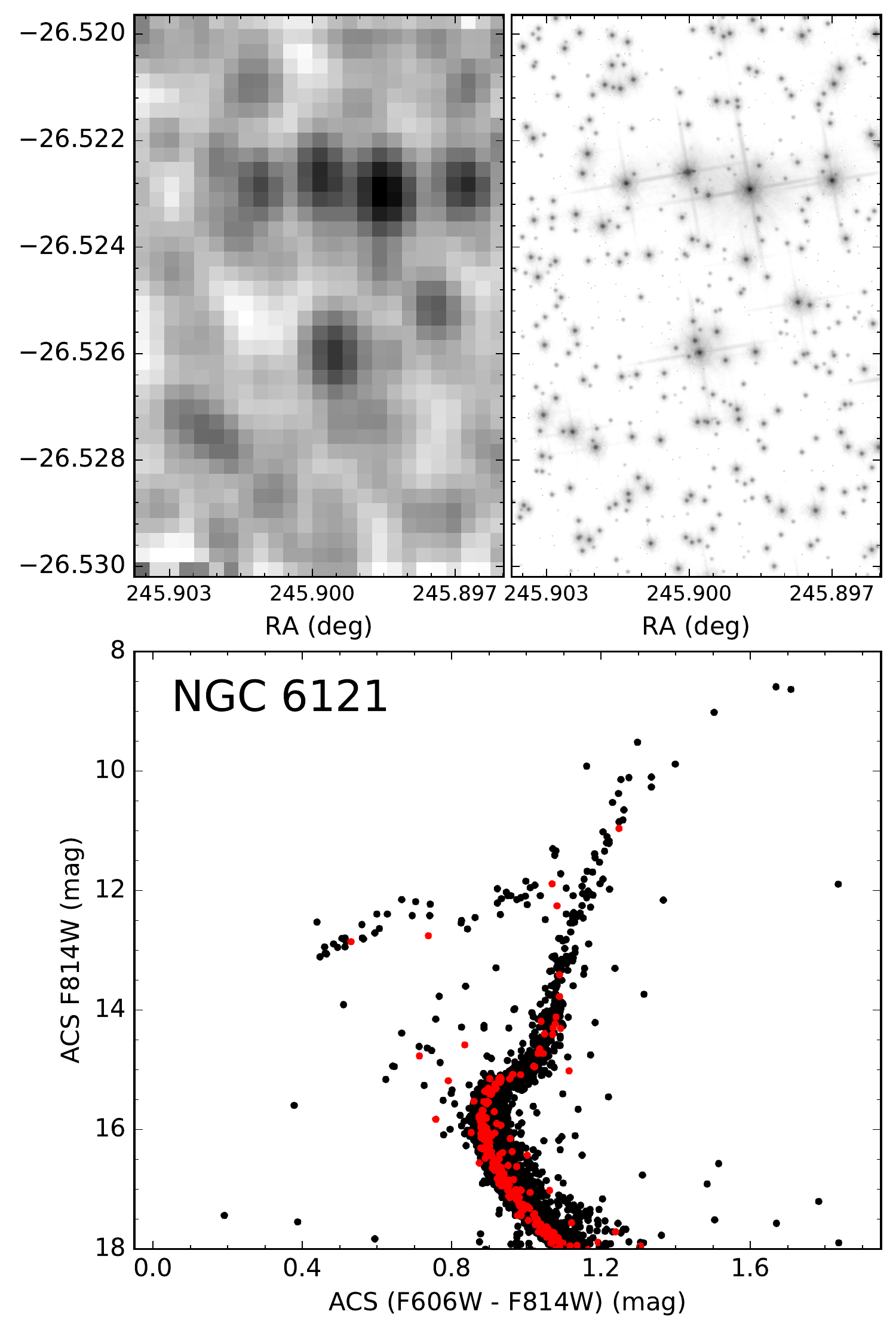}
\caption{Comparison of the NGC 6121 WiFeS datacube with ACS imaging.
\emph{Top left:} WiFeS I7000 datacube collapsed along the wavelength direction.
\emph{Top right:} HST/ACS Wide Field Camera F814W image (GO-10775) of the same field-of-view as the WiFeS datacube.
\emph{Bottom:} ACS Globular Cluster Treasury Survey \citep{2008AJ....135.2055A} (F606W $-$ F814W) colour-magnitude diagram.
Black points are all stars in the ACS catalogue while red points are stars in the field-of-view of the WiFeS datacube.
Our NGC 6121 datacube poorly samples the GC's colour-magnitude diagram with only a handful of red giant branch and horizontal branch stars in the field-of-view.
The brightest star in the field-of-view is a F814W$ = 11$ red giant star while faintest stars in the datacube that are not blends are subgiant branch stars with F814W$ \sim 15$.
We note NGC 6121 is the closest GC in our sample and has the fourth lowest luminosity calculated enclosed by the field-of-view.}
\label{fig:NGC6121_acs_waggs}
\end{center}
\end{figure}

\begin{figure}
\begin{center}
\includegraphics[width=240pt]{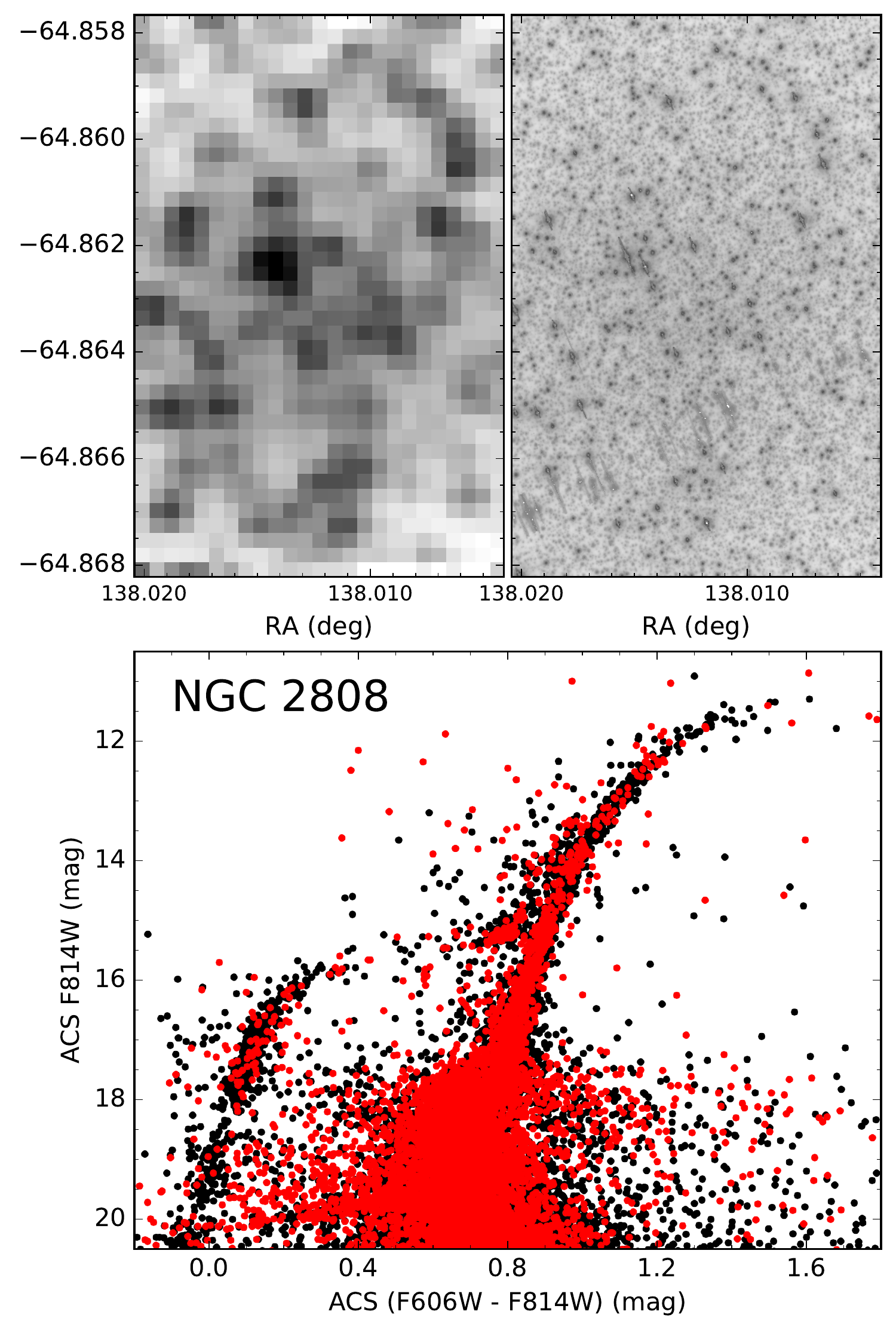}
\caption{Comparison of the NGC 2808 WiFeS datacube with ACS imaging.
\emph{Top left:} WiFeS I7000 datacube collapsed along the wavelength direction.
\emph{Top right:} HST/ACS Wide Field Camera F814W image (GO-10775) of the same field-of-view as the WiFeS datacube.
\emph{Bottom:} ACS Globular Cluster Treasury Survey \citep{2008AJ....135.2055A} (F606W $-$ F814W) colour-magnitude diagram.
Black points are all stars in the ACS catalogue while red points are stars in the field-of-view of the WiFeS datacube.
The HST colour-magnitude diagram is well sampled by our observations.
Our spatial coverage of NGC 2808 is typical of our sample although it is one of the more massive GCs in our sample.}
\label{fig:NGC2808_acs_waggs}
\end{center}
\end{figure}

As can be seen in Figure \ref{fig:field_of_view}, individual stars can be seen in the datacubes of the closest or lowest surface brightness GCs.
In Figures \ref{fig:NGC6121_acs_waggs} and \ref{fig:NGC2808_acs_waggs} we compare our WiFeS datacubes with HST/ACS imaging from the ACS Globular Cluster Treasury Survey \citep{2007AJ....133.1658S}, downloaded from the Hubble Legacy Archive \footnote{\url{http://hla.stsci.edu/}}.
In each of these figures we also show the colour-magnitude diagrams (F606W $\sim V$, F814W $\sim I$) for all stars in the ACS Globular Cluster Treasury Survey catalogues \citep{2008AJ....135.2055A} and those stars within the WiFeS field-of-view.
For NGC 6121, the brightest star in the WiFeS field-of-view is a F814W$ = 10.97$ red giant while the faintest identifiable stars are F814W$ \sim 15$ subgiant branch stars.
We note that many of the 'stars' in the WiFeS cubes are in fact blends of multiple stars.
As can be seen in the lower panel of Figure \ref{fig:NGC6121_acs_waggs}, the colour-magnitude diagram of NGC 6161 is not well sampled with only one red giant branch star brighter than the horizontal branch in the WiFeS field-of-view. 
The colour-magnitude diagram of NGC 2808 (Figure \ref{fig:NGC2808_acs_waggs}), however, is well sampled by the WiFeS field-of-view.
We note that NGC 2808 is close to the median of our sample in terms of distance from the Sun and at the eightieth percentile of the $V$-band luminosity enclosed by the WiFeS field-of-view while NGC 6121 is the closest GC in our sample and has the fourth lowest enclosed luminosity.

\begin{figure}
\begin{center}
\includegraphics[width=240pt]{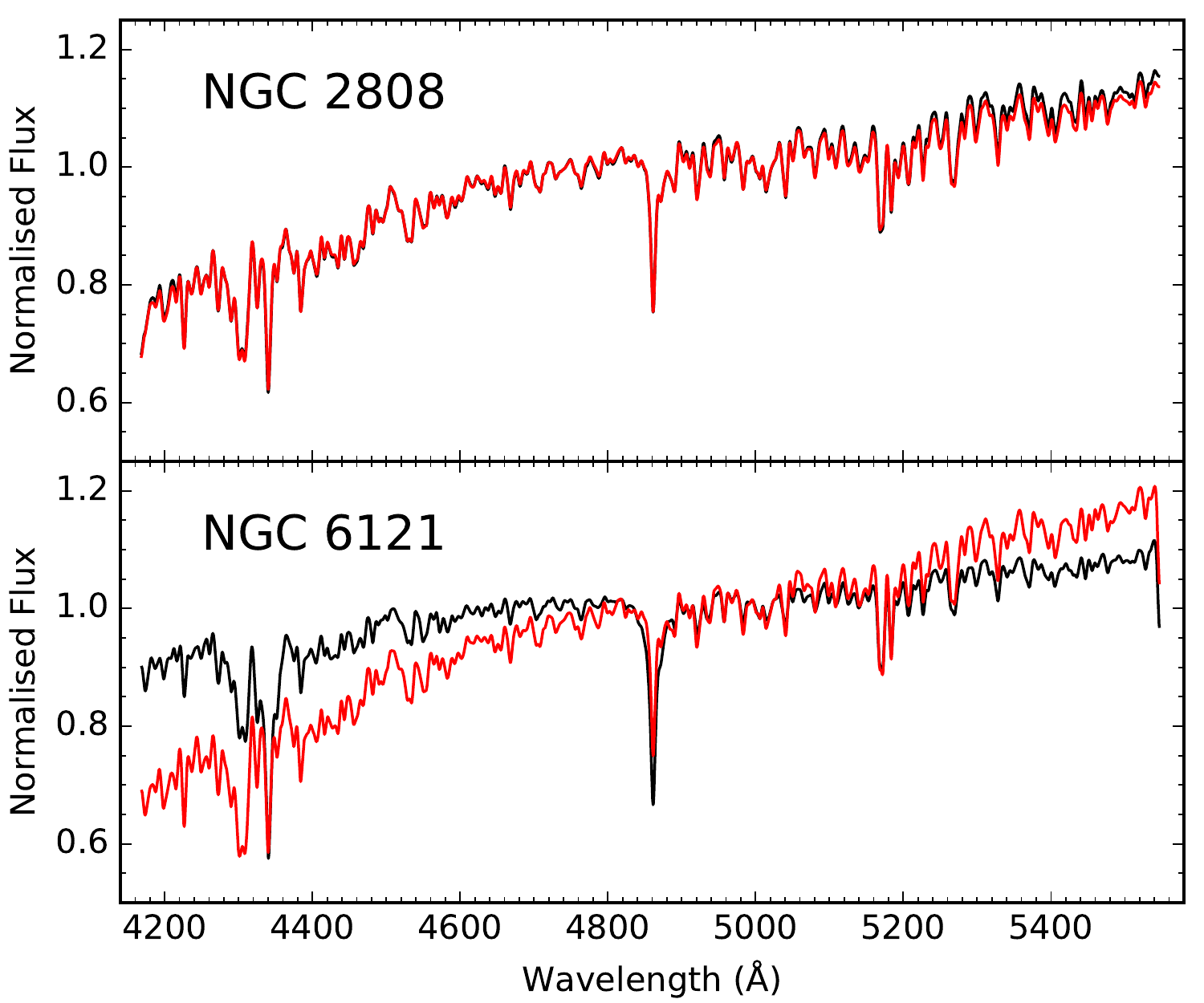}
\caption{Effects of stochastic fluctuations in the number of stars within the field-of-view on the integrated spectra.
In both panels the spectrum extracted from the left-hand side of the B7000 cube is shown in black and the spectrum from the right in red.
For the purposes of clarity, the spectra have been shifted to the rest frame,  smoothed by a 5 \AA{} FWHM Gaussian and normalised such that the mean flux at 5000 \AA{} is unity.
In the case of NGC 6121, where the observed light is dominated by a handful of red giant branch and horizontal branch stars, there are large differences the shape of the continuum and the strengths of spectral features such as H$\gamma$ and H$\beta$ in the spectra from the two halves of the datacube.
In the more typical case of NGC 2808, there are only minor but significant differences between the two halves.}
\label{fig:half_test}
\end{center}
\end{figure}

To assess what effect this stochastic sampling of the colour-magnitude diagrams has on the integrated spectra, we divided our datacubes of NGC 2808 and NGC 6121 each in two and extract spectra for each half cube.
As can be seen in Figure \ref{fig:half_test}, the differences between the two halves of the datacube are large for NGC 6121 while for NGC 2808, the differences are small but significant.
The effects of stochastic sampling are largest for the closest GCs and for the GCs with lowest enclosed luminosity. 
Stochastic sampling of the IMF is also an issue for the younger LMC and SMC GCs as their integrated light can be dominated by a handful of supergiants or thermally pulsing asymptotic giant stars.
The effects of stochastic sampling are generally larger than the statistical uncertainty of the integrated spectra but are wavelength and spectral feature dependent, with spectral features produced by rare but luminous stars, such as TiO bands produced by cool giants, showing the largest variability.

\subsection{First data release}
The integrated spectra of the WAGGS sample are available at \url{http://www.aao.gov.au/surveys/waggs}.
We note that in this first data release spectra have been flux calibrated only in a relative sense.
Future data releases will include the full datacubes, improved flux calibration and improved telluric corrections.
We also aim to enlarge our sample with further observations of MW, LMC and SMC GCs.

\section{Summary and future work}
\label{sec:summary}
We have presented WAGGS, a new library of integrated GC spectra.
We have used the WiFeS integral field spectrograph to observe the centres of 64 MW, 14 LMC, 5 SMC and 3 Fornax dSph GCs. 
As can be seen in the lower left panel of Figure \ref{fig:sample_properties} and Figures \ref{fig:blue_metal} to \ref{fig:red_age}, our sample spans a wide range of metallicities ($-2.4 <$ [Fe/H] $< - 0.1$) and ages (20 Myr to 13.5 Gyr).
The WAGGS spectra have significantly higher spectral resolution ($R = 6800$) compared to earlier studies (Figure \ref{fig:effect_resolution}) while Figure \ref{fig:ngc104} demonstrates the wide wavelength coverage (3300 to 9050 \AA) of our spectra. 

The WAGGS dataset will be quite useful for a range of applications in Galactic and extra-galactic astronomy.
The most obvious use is testing stellar population synthesis models.
We note that the spectral resolution and wavelength coverage of WAGGS spectra exceed those of the commonly used MILES \citep{2006MNRAS.371..703S} empirical stellar library. 
The spatially resolved nature of the WAGGS datacubes makes possible for the stars contributing to the integrated spectra to be identified in resolved imaging.
This allows the luminosity function from high spatial resolution imaging of the same area on the sky (for example HST based photometry) to be used in place of an assumed IMF.
The WAGGS spectra enable comparisons of measurements of IMF sensitive spectral features such as the sodium doublet at 8190 \AA{} and the CaT with models.
The effects of GC dynamical evolution, namely mass segregation, could be used to provide stellar populations with the same ages and chemical compositions but different present day mass functions.

The dataset will also makes possible the testing of a range of stellar population analysis techniques including spectral indices \citep[e.g.][]{1994ApJS...95..107W, 2006MNRAS.370.1106G, 2007ApJS..171..146S, 2008ApJS..177..446G}, full spectral fitting \citep[e.g.][]{2008MNRAS.385.1998K, 2014ApJ...780...33C} and spectral synthesis of narrow spectral regions \citep[e.g.][]{2009ApJ...704..385C, 2012A&A...546A..53L, 2013MNRAS.434..358S} as well as hybrid techniques \citep[e.g. template based measurements of the CaT,][Usher et al. in prep.]{2010AJ....139.1566F, 2012MNRAS.426.1475U}.
The semi-resolved nature of many of the datacubes can be used to test novel semi-resolved stellar population analysis techniques \citep[e.g.][]{2014ApJ...797...56V, 2016ApJ...827....9C}.
WAGGS spectra can also be used to derive empirical relations between spectral indices and parameters such as metallicity \citep[e.g.][]{1991ApJ...379..157B, 2004AJ....128.1671S, 2010AJ....140.2101S}, horizontal branch morphology \citep[e.g.][]{2004ApJ...608L..33S} or blue straggler frequency \citep{2008ApJ...689L..29C}.
The Local Group GCs in WAGGS can now be compared in a model independent way with observations of extragalactic GCs.

The WAGGS spectra will also be a boon to studies of the GCs of the MW and its satellites.
Our current sample covers 40 per cent of the MW's GC system.
The high S/N, intermediate resolution spectra may be used to generate a homogeneous abundance scale based on a large number of GCs, both in the MW and its satellites.
Since the velocity resolution of the WiFeS spectra ($\sigma = 19$ km s$^{-1}$) is comparable to the central velocity dispersion of the more massive GCs, the WAGGS datacubes should provide useful constraints to dynamical modelling.
Improved dynamical GC masses at high metallicities and at younger ages would allow better understanding of how the M/L varies with age and metallicity.
In future we plan on exploring the use of point spread function fitting to extract spectra of individual stars from our datacubes in a manner analogous to point spread function photometry \citep[e.g.][]{2004ApJ...603..531R, 2013A&A...549A..71K, 2013MNRAS.430.1219P, 2016A&A...588A.148H}.

The WAGGS spectra will be extremely useful to test stellar population synthesis models and analysis techniques, to compare with extragalactic GC observations and to study the sample GCs themselves.
We have made the first data release of the integrated spectra publicly available on the project website.
Future papers in this series will focus on using the WAGGS dataset to address a number of scientific questions. 

\section*{Acknowledgements}
The authors wish to thank the referee S\o ren Larsen for his useful comments and suggestions which helped to improve this paper. 
We wish to thank Nate Bastian, Joel Pfeffer and Thomas Beckwith for helpful discussions and useful suggestions.
We are grateful to Paolo Bonfini, Martina Fagioli, Bililign Dullo, Elisa Boera, Srdan Kotus and Paul Frederic Robert for their assistance with the observations.
We wish to thank Simon O'Toole for his assistence with hosting the first WAGGS data release.
We also wish to greatly thank the Siding Spring Observatory staff for their assistance with the ANU 2.3 m telescope and WiFeS instrument.
We also wish to thank Jeremy Mould for originally suggesting the use of WiFeS to obtain integrated spectra of Local Group globular clusters.
C.U. wishes to thank the University of Victoria's Department of Physics and Astronomy for their hospitality while a portion of this work was carried out.
C.U. gratefully acknowledges financial support from the European Research Council (ERC-CoG-646928, Multi-Pop).
P.C. acknowledges the support provided by FONDECYT postdoctoral research grant no 3160375 and by the Swinburne Chancellor Research Scholarship and the AAO PhD Topup Scholarship during the 2015 observing runs.
S.B. acknowledges the support of the AAO PhD Topup Scholarship.
This work made use of \textsc{NumPy} \citep{numpy}, \textsc{Scipy} \citep{scipy}, and \textsc{matplotlib} \citep{Matplotlib} as well as \textsc{Astropy}, a community-developed core Python package for astronomy \citep{2013A&A...558A..33A}.
Additionally, this work made use of \textsc{TOPCAT} \citep{2005ASPC..347...29T} and \textsc{Aladin} \citep{2000A&AS..143...33B}.
This work was partially performed on the swinSTAR supercomputer at Swinburne University of Technology.
This publication makes use of data products from the Two Micron All Sky Survey, which is a joint project of the University of Massachusetts and the Infrared Processing and Analysis Center/California Institute of Technology, funded by the National Aeronautics and Space Administration and the National Science Foundation.
This research made use of Montage, funded by the National Aeronautics and Space Administration's Earth Science Technology Office, Computational Technnologies Project, under Cooperative Agreement Number NCC5-626 between NASA and the California Institute of Technology. The code is maintained by the NASA/IPAC Infrared Science Archive.
This research has made use of the NASA/IPAC Extragalactic Database (NED), which is operated by the Jet Propulsion Laboratory, California Institute of Technology, under contract with the National Aeronautics and Space Administration.
This researched is based on observations made with the NASA/ESA Hubble Space Telescope, and obtained from the Hubble Legacy Archive, which is a collaboration between the Space Telescope Science Institute (STScI/NASA), the Space Telescope European Coordinating Facility (ST-ECF/ESA) and the Canadian Astronomy Data Centre (CADC/NRC/CSA). 

\bibliographystyle{mnras}
\bibliography{bib}{}

\end{document}